# Rates of protoplanetary accretion and differentiation set nitrogen budget of rocky planets

Damanveer S. Grewal[1]*, Rajdeep Dasgupta[1], Taylor Hough[1], Alexandra Farnell[1,2]

[1]Department of Earth, Environmental, and Planetary Sciences, Rice University, 6100 Main Street, MS 126, Houston, TX 77005, USA
[2]St. John's School, 2401 Claremont Ln, Houston, TX 77019, USA

*correspondence: dsg10@rice.edu

**The effect of protoplanetary differentiation on the fate of life-essential volatiles like nitrogen and carbon and its subsequent effect on the dynamics of planetary growth is unknown. Because the dissolution of nitrogen in magma oceans depends on its partial pressure and oxygen fugacity, it is an ideal proxy to track volatile re-distribution in protoplanets as a function of their sizes and growth zones. Using high pressure-temperature experiments in graphite-undersaturated conditions, here we show that the siderophile (iron-loving) character of nitrogen is an order of magnitude higher than previous estimates across a wide range of oxygen fugacity. The experimental data combined with metal-silicate-atmosphere fractionation models suggest that asteroid-sized protoplanets, and planetary embryos that grew from them, were nitrogen-depleted. However, protoplanets that grew to planetary embryo-size before undergoing differentiation had nitrogen-rich cores and nitrogen-poor silicate reservoirs. Bulk silicate reservoirs of large Earth-like planets attained nitrogen from the cores of latter type of

**planetary embryos. Therefore, to satisfy the volatile budgets of Earth-like planets during the main stage of their growth, the timescales of planetary embryo accretion had to be shorter than their differentiation timescales, i.e., Moon- to Mars-sized planetary embryos grew rapidly within ~1-2 Myrs of the Solar System's formation.**

Bulk silicate reservoirs (mantle + crust + atmosphere) of rocky protoplanets and planets in the inner Solar System are extremely depleted in major volatiles like nitrogen (N) and carbon (C)[1–5]. It is assumed that these bodies grew by accreting nearly volatile-free material due to a lack of condensation of volatiles in their growth zone[2,6]. Therefore, volatile inventory of rocky planets is thought to be chiefly established by the addition of carbonaceous chondrite-like material from the outer Solar System to almost volatile-free bodies[1,2]. However, enstatite chondrites that source the inner Solar System reservoir[7] contain several hundreds of ppm of N and C in refractory phases[8]. The presence of isotopically distinct N in NC (non-carbonaceous chondrite affinity) iron meteorites also suggests that planetesimals and planetary embryos in the inner Solar System did not accrete volatile-poor material[9]. Consequently, post-accretion processes should have played an important role in establishing the N- and C-depleted character of the bulk silicate reservoirs of rocky bodies.

Early differentiation processes such as core-mantle separation and atmospheric loss play an important role in explaining N- and C-depletion in the bulk silicate Earth (BSE)[3,4,10–15]. However, the effect of a differentiated character of the earliest forming protoplanets on the dynamics of N- and C-retention and loss during planetary growth is not known. Several lines of evidence point towards a differentiated character of these protoplanets independent of their growth zone: 1) rapid accretion and differentiation of asteroid- and planetary embryo-sized rocky bodies[16] – including parent bodies of CC (carbonaceous chondrite affinity) and NC iron meteorites[17] – when $^{26}$Al was extant; 2) geochemical evidence for the presence of global magma oceans (MOs) on asteroid-sized Vesta and angrite parent bodies[18]; and 3) paleomagnetic evidence of magnetic core dynamos in the parent bodies of CV and CM carbonaceous chondrites which originated in the outer Solar System[19,20]. If differentiation was ubiquitous in the earliest forming protoplanets at varying heliocentric distances, can it explain the widespread N- and C-depleted character of the bulk silicate reservoirs of all rocky bodies independent of their sizes and timescales of their growth?

In the earliest formed protoplanets, large-scale melting by $^{26}$Al decay triggered metal-silicate separation as well as volatile degassing[21,22]. Vapor pressure-imposed solubility determines volatile distribution between the overlying proto-atmosphere and silicate MO, and consequently, the amount of volatiles partitioned into the core[3,14,23]. Unlike C, whose concentration in MOs is challenging to determine due to the formation of accessory phases like graphite[14,23], N is an ideal proxy to explore the fate of volatiles during protoplanetary differentiation. This is because the vapor pressure-imposed solubility of N in the silicate melts is controlled by $fO_2$ and $p_N$[24], and $D_N^{alloy/silicate}$ (concentration of N in alloy/concentration of N in silicate) by $fO_2$[11–13], with much smaller effects of other thermodynamic parameters. $D_N^{alloy/silicate}$ combined with vapor pressure-imposed N solubility in the silicate melts can thus be used to constrain coupled N partitioning between all three reservoirs, accounting for the variation in the composition of accreting material as well as the sizes of the protoplanets (Fig.1).

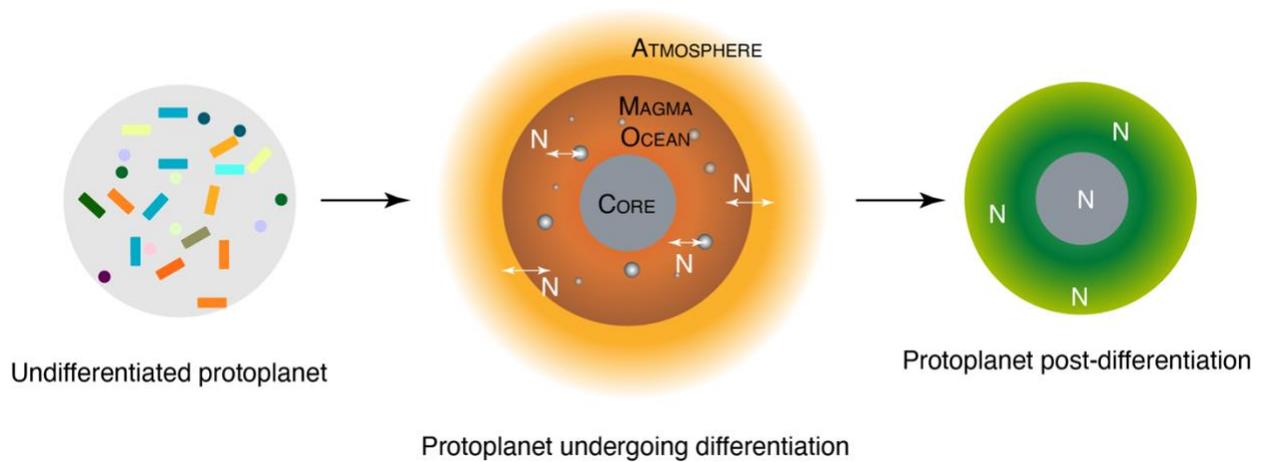

*Figure 1: An illustration showing the fate of nitrogen during protoplanetary differentiation. An undifferentiated protoplanet represents an amalgamation of primitive chondrite-like material. Heat releasing during the decay of $^{26}$Al causes large-scale melting which triggers metal-silicate separation as well as volatile degassing[21,22]. Vapor-pressure induced solubility sets N abundances in the overlying atmosphere and magma ocean, and consequently the amount of N partitioned into the core via equilibrium exchange between magma ocean-core-atmosphere reservoirs. Post-differentiation, N retained in the silicate and metallic reservoirs only is available during subsequent stages of planetary growth while the proto atmospheres are lost*

*either due to the inability of the smaller bodies to retain their atmospheres or due to impact-induced erosion.*

**Partitioning of nitrogen between alloy and silicate melts**

$D_N^{alloy/silicate}$ as a function of $fO_2$ has chiefly been determined in graphite-saturated conditions yielding C-rich alloys containing 3.5-5.5 wt.% C[10–13,25]. Experiments at ambient[26] and high pressure[27] have shown that C has a strong negative effect on N dissolution in Fe, Ni-alloys because N and C occupy similar interstitial voids in the alloy melt structure. As typical core-forming alloys are not expected to be graphite-saturated[14], previous experimental studies might have underestimated $D_N^{alloy/silicate}$ if similar negative interactions between N and C persist for *P-T-fO₂* conditions relevant for protoplanetary and planetary core-mantle differentiation. Some data from ref.[15] in a limited $fO_2$ range support this hypothesis but remains unquantifiable (see Methods). It is unknown whether the effect of C content in alloy on $D_N^{alloy/silicate}$ persists across a wide $fO_2$ range applicable for protoplanetary and planetary differentiation.

To constrain $D_N^{alloy/silicate}$ in graphite-undersaturated conditions, we performed piston cylinder experiments using Fe-Ni-N±Si alloys and silicate mixtures in MgO capsules at 3 GPa and 1600-1800 °C (Supplementary table 1). The rationale for the chosen experimental conditions is based on previous studies[10–13,25] which showed that $D_N^{alloy/silicate}$ is primarily controlled by $fO_2$ with *P* having a minimal effect for alloy-silicate equilibration in planetesimals and planetary embryos[11,12,15]. Reaction of silicate mixtures with MgO capsules generated ultramafic silicate melts that better represent planetary MO compositions (Extended Data Fig.1). To cover the compositional range for protoplanets accreted at disparate heliocentric distances, the experiments covered a wider $fO_2$ range (IW–7.10 to –1.54; IW refers to $fO_2$ set by equilibrium coexistence of iron (Fe) and wüstite (FeO); Supplementary table 2) in comparison to previous studies (IW–4.20 to –0.07; refs.[10–13,15,25]). A time series conducted between 0.5 and 12 hours demonstrates minimal variations in N contents in alloy and silicate melts as well as silicate melt compositions with time (Extended Data Fig.2; see Methods). Also, there was no zonation of N content in either alloy melt or silicate phases. Collectively these demonstrate that the experiments had reached equilibrium at less than 0.5 hours. Therefore, a runtime of 45 minutes to 3 hours – comparable to previous studies in graphite-saturated conditions[10–13] – was deemed sufficient to determine equilibrium $D_N^{alloy/silicate}$ values. Experimental products comprised metal blobs embedded either in silicate glass (Fig.2a) or in a matrix of silicate glassy pools along with quenched dendritic crystals (Fig.2b), fine quenched crystals (Fig.2c), or euhedral olivine (Fig.2d), and periclase. N and other major and minor element abundances in alloy and glassy melts were determined using electron microprobe (Supplementary table 3,4; see Methods). Only N-H peak was spectroscopically observed in the silicate glasses (Extended Data Fig.3; see Methods). In experiments comprising both glassy and quenched crystal domains, N contents in the glassy pools was used to estimate N concentration of the silicate melt because dendritic matte formed during quenching loses N due to melt compaction[11].

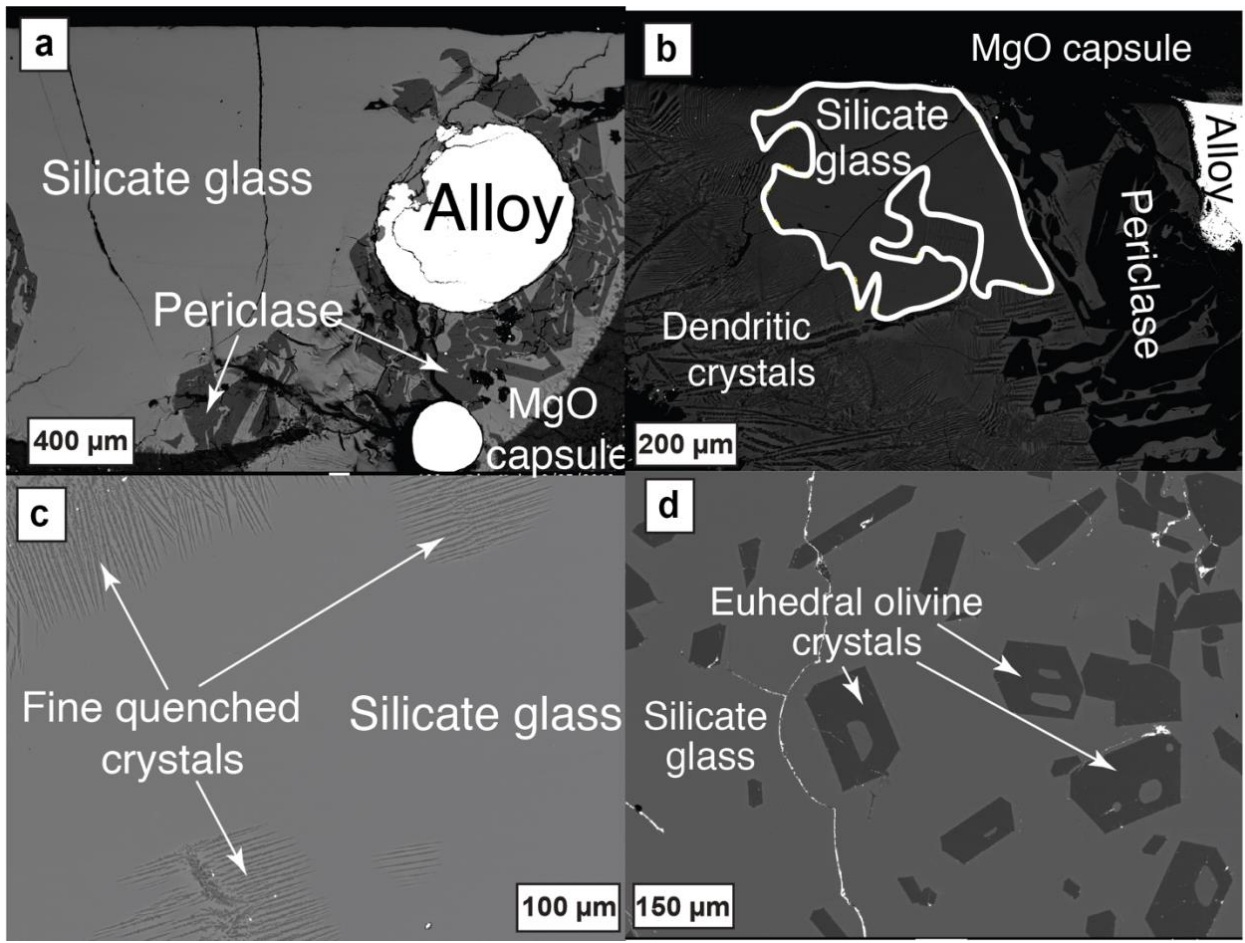

*Figure 2: Back-scattered electron images of experimental products: (a) Co-existence of quenched alloy blobs and silicate glass with periclase in MgO capsule (G608; 3 GPa-1700 ºC-IW–6.62). (b) Co-existence of quenched alloy blobs with silicate glass pools, dendritic olivine crystals, and periclase (X25; 3 GPa-1600 ºC-IW–3.51). (c) Texture of silicate glass co-existing*

*with fine quenched crystals (X39; 3 GPa-1700 ºC-IW–4.60). **(d)** Texture of silicate glass co-existing with euhedral olivine crystals (X43; 3 GPa-1800 ºC-IW–4.22).*

C contents in graphite-undersaturated alloy melts of this study vary between 0.11 and 0.80 wt.% and are substantially lower than graphite-saturated alloys of previous studies (3.5-5.5 wt.%)[10–13] (Extended Data Fig.4a). Even though the starting mixtures were nominally C-free, atmospheric contamination during preparation and/or diffusion of carbon from the graphite heater during the experiments could have been the potential source of C in the alloy melts. In agreement with graphite-saturated experiments[11], N content in the alloy decreases with decreasing $fO_2$ followed by a sharp drop at ~IW–4, caused by incorporation of Si in the alloy melt (Extended Data Fig.4b). This drop results from an increase in $\gamma_N^{alloy\ melt}$ (activity coefficient of N in the alloy melt with infinitely dilute solution as the standard state) with increasing Si due to non-ideal interactions between Si and N (Fig.3a). At a given $fO_2$, N contents are higher in graphite-undersaturated alloys by a factor of ~2-10 relative to graphite-saturated alloys (Extended Data Fig.4b). $\gamma_N^{alloy\ melt}$ for Si-free alloys increases from ~1 for graphite-undersaturated alloys to ~3-4 for graphite-saturated alloys at a fixed $fO_2$ (Fig.3a). This confirms that negative interactions between N and C also persist at high *P-T* relevant for protoplanetary differentiation and indicates N dissolution in core forming alloys melts was severely underestimated while using graphite-saturated alloys. Consequently, at any given $fO_2$, $D_N^{alloy/silicate}$ for graphite-undersaturated alloys is nearly an order of magnitude higher than graphite-saturated alloys (Fig.3b). N acts as a siderophile element over a wider $fO_2$ range (>~IW–4.5) in graphite-undersaturated conditions relative to graphite-saturated conditions (>~IW–2). The decrease in $D_N^{alloy/silicate}$ with decreasing $fO_2$ is due to an increase in $\gamma_N^{alloy\ melt}$ and stability of $N^{3-}$ in the silicate melt[24,28] under extremely reducing conditions. Also, we find that at a similar log$fO_2$ (~IW–4), incorporation of Si into the alloy melt decreases $D_N^{alloy/silicate}$ substantially (Extended Data Fig.5a). In contrast with previous studies, $D_N^{alloy/silicate}$ does not vary with *T* (Extended Data Fig.5b). ~50-85 % of N was recovered in the final products (Extended Data Fig.6). This is within the range of or higher than N recovery in graphite capsules[10–13,25,29]. Lack of a correlation between N loss and $D_N^{alloy/silicate}$ and correlations of $D_N^{alloy/silicate}$ with other thermodynamic parameters (e.g., $fO_2$ and C content of alloy) suggests adherence to Henry's law.

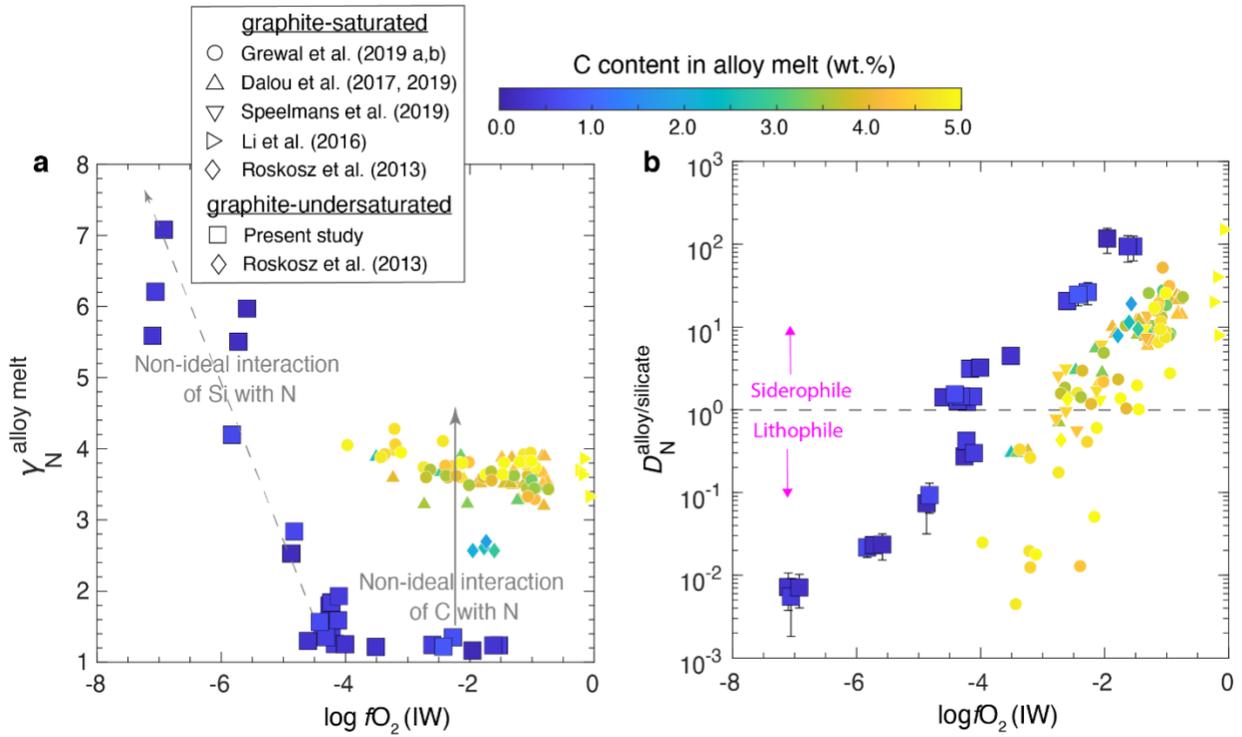

*Figure 3: $\gamma_N^{alloy\ melt}$ and $D_N^{alloy/silicate}$ as a function of oxygen fugacity and carbon content in the alloy melt. a) $\gamma_N^{alloy\ melt}$ in C-poor and Si-free alloy melts in this study is ~1 while $\gamma_N^{alloy\ melt}$ is ~3-4 at similar $fO_2$ for graphite-saturated alloys. This is due to non-ideal interactions between C and N occupying similar octahedral voids in the alloy melt. Higher $\gamma_N^{alloy\ melt}$ for C-poor and Si-bearing alloys results from repulsive interactions between Si and N. b) In agreement with previous studies in graphite-saturated conditions[10–13,25,29], $D_N^{alloy/silicate}$ decreases with decreasing $fO_2$ in graphite-undersaturated conditions. At any given $fO_2$, $D_N^{alloy/silicate}$ in graphite-undersaturated conditions is almost an order of magnitude higher relative to graphite-saturated conditions. $\gamma_N^{alloy\ melt}$ was calculated using the 'Online Metal Activity Calculator' (http://norris.org.au/expet/metalact/) which uses $\varepsilon$ approach via Wagner equations. Error bars for $D_N^{alloy/silicate}$ represent ±1-σ deviation obtained by propagation of ±1-σ deviation on N content in the alloy and silicate melts; where absent, the error bars are smaller than the symbol size.*

**Alloy-silicate-atmosphere fractionation of nitrogen**

The fate of N during protoplanetary differentiation depends on the growth rate and final sizes of protoplanets that formed within ~1-2 Myr of CAI formation as well as their chemical compositions (tracked here by $fO_2$). Numerical and cosmochemical models predict an almost instantaneous accretion of Vesta-sized asteroids and Moon- to Mars-sized planetary embryos within ~0.1-2 Myr of CAI formation[16,30–32]. Even though smaller planetesimals were more numerous, most of the mass was distributed in asteroid- and planetary embryo-sized bodies[30,31]. To account for variation in sizes of protoplanets, we assume four end-members cases – Vesta (R~250 km; 0.04 $R_⊕$; 0.00004 $M_⊕$), an intermediate-sized hypothetical protoplanet (R~750 km; 0.12 $R_⊕$; 0.001 $M_⊕$), Moon (R~1740 km; 0.27 $R_⊕$; 0.012 $M_⊕$), and Mars (R~2440 km; 0.53 $R_⊕$; 0.107 $M_⊕$; $R_⊕$ and $M_⊕$ is the radius and mass of present-day Earth, respectively). Once a rocky body grew rapidly to its final size[16,30,31], it is assumed to have undergone global melting leading to the formation of an MO (via heat released by $^{26}Al$ decay) overlain by a degassed atmosphere[21,22] (Fig.1). Vapor pressure induced solubility sets the N content in the MO which in turn determines the availability of N to partition into core-forming alloy. Alloy-silicate equilibration sets the $fO_2$ of the MO. The effect of compositional gradient of the accreting material is accounted by alloy-silicate equilibration at different $fO_2$s. It is important to note that differentiation at log $fO_2$> IW–4 is representative for most rocky bodies in the Solar System and only Mercury and aubrite parent body underwent differentiation at more reducing conditions (log $fO_2$< IW–5)[33] (Fig.4b). As alloy-silicate equilibration was extremely efficient in planetesimal- and planetary embryo-sized bodies[34], complete alloy-silicate melt equilibration was assumed. N abundances in the atmosphere, MO, and alloy melts are calculated simultaneously using parametrized N solubility equation from ref.[24] and parameterized $D_N^{alloy/silicate}$ equation form this study (Eq.3; Supplementary table 5; Extended Data Fig.7).

Distribution of N between the atmosphere, MO, and core reservoirs as a function of $fO_2$ for end-member protoplanets is shown in Fig.4 for a fixed alloy/silicate mass ratio (r = 0.5; Earth's alloy/silicate ratio), bulk N content of 500 ppm (within the range of carbonaceous and enstatite chondrites[8]), and alloys having 0-0.4 wt.% C (estimated upper limit of C content of Earth's core[4]). The distribution of N within different reservoirs depends on a complex interplay between N solubility in MO and $D_N^{alloy/silicate}$, which directly depends on the size of the body and $fO_2$ of the accreting material. At any given $fO_2$, percentage of N in the atmospheric reservoir decreases with increase in the size of the rocky body (Fig.4a). Therefore, a Vesta-sized protoplanet has almost all of its N inventory in the atmospheric reservoir (except below IW–5), while larger protoplanets have proportionally lesser N (Fig.4a). Assuming density of the material accreted by protoplanets was similar, $p_N$ scales as function of radius (R) and gravitational constant (g) such that larger bodies have higher $p_N$ at a given $fO_2$ (Fig.4d). Hence, more N dissolves into the MOs of larger bodies at a fixed $fO_2$ (Fig.4b). Consequently, a greater amount of N is available for fractionation between MO and alloy melts, allowing a higher proportion of N to segregate into the cores of larger bodies at a given $fO_2$ (Fig.4c). For all rocky bodies, N in the atmospheric reservoir increases from IW–7 to IW–4 because N solubility in the MO drops with increase in $fO_2$ and the core cannot incorporate substantial amount of N due to $D_N^{alloy/silicate}$ being lower than 1 under reducing conditions. Above IW–4, N in the atmospheric reservoir decreases (especially for larger bodies) even though its solubility in the MO decreases because N shows an increasingly siderophile character with increasing $fO_2$. Accordingly, cores become an important reservoir for N storage in larger protoplanets. For example, cores contain up to ~40% and ~70% of the accreted N budget in Moon- and Mars-sized planetary embryos, respectively. As a result, MOs relevant only for Mercury and aubrite parent body (log$fO_2$ <IW–4) retain a substantial fraction of N in their silicate reservoirs. Above log$fO_2$> IW–4, MOs of protoplanets are extremely N depleted (<0.1% of the initial accreted N) irrespective of their sizes – either solely due to degassing of N into the atmosphere or in combination with N segregating into the cores. Although Moon- to Mars-sized planetary embryos allow vapor pressure buildup leading to substantial volatile ingassing into MOs and cores[21,35], such vapor pressure build up is unlikely for asteroid-sized bodies due to rapid atmospheric escape as a result of their small gravitational pull[21,22]. Therefore, our predictions for N in the MOs and cores for Vesta- and intermediate-sized bodies represent an upper limit.

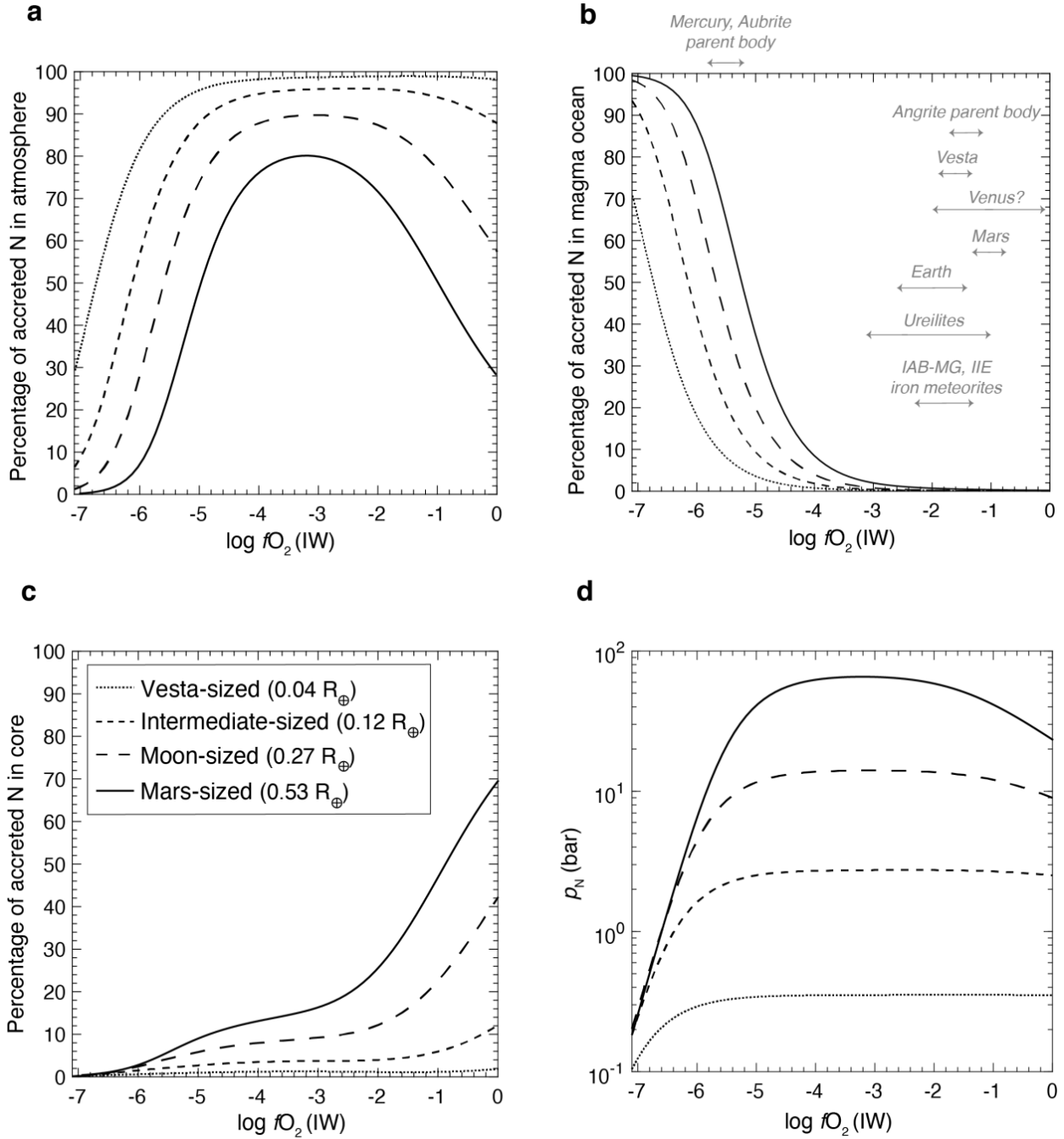

*Figure 4: Relative distribution of nitrogen in the constituent reservoirs of protoplanetary bodies as a function of oxygen fugacity. At any given fO2, the proportion of N in the (**a**) atmospheric reservoir decreases, while it increases in the (**b**) magma ocean and (**c**) core reservoir with increase in the size of the rocky body. Assuming these bodies had similar densities, their N partial pressure (pN) correlate with their radius. Therefore, large bodies have higher pN (**d**) resulting in a higher amount of N dissolving into their MOs and cores. As N*

*solubility in the MO decreases and the siderophile character of N increases with increasing $fO_2$, the proportion of N in the MO decreases and proportion of N in the atmosphere and/or core increases. Consequently, for rocky body differentiation at $\log fO_2 >$ IW–4, asteroid-sized (Vesta (0.04 $R_\oplus$) and intermediate-sized (0.12 $R_\oplus$)) bodies have almost all of their N in the atmospheric reservoir, while for planetary embryo-sized bodies (Moon (0.27 $R_\oplus$) and Mars (0.53 $R_\oplus$)) the cores become an important reservoir for N storage. The $fO_2$s of core-mantle differentiation in panel (**b**) are compiled in Supplementary table 6.*

Although we assumed fixed bulk N and alloy/silicate mass ratio in Fig.4, the relative distribution of N during differentiation remains practically unchanged irrespective of the variations in the alloy/silicate mass ratio or the efficiency of alloy-silicate equilibration (Supplementary Fig.1) and initial amount of accreted N (Supplementary Fig.2). This suggests that the variation in size of the protoplanets and the composition of their accreting material control the relative distribution of N during protoplanetary differentiation. However, using $D_N^{\text{alloy/silicate}}$ determined for graphite-undersaturated alloys is critical for the conclusions drawn in Fig.4. $D_N^{\text{alloy/silicate}}$ is distinctly lower for graphite-saturated systems which results in atmosphere being the dominant N-bearing reservoir at $\log fO_2 >$ IW–4, irrespective of the size of the parent body (Supplementary Fig.3).

**Rate of protoplanetary accretion versus differentiation**

The earliest formed planetesimals and planetary embryos via instantaneous accretion act as seeds for the planetary embryos and planets that grew later. Therefore, the relative distribution of N between their atmospheric and non-atmospheric reservoirs has important implications for the fate of volatiles during the next stage of planetary growth via planetesimal and planetary embryo collisions. High stellar heating during the protoplanetary disk stage and low escape velocities of the earliest-forming asteroids and planetary embryos makes it improbable that the earliest formed protoplanets retained their MO degassed atmospheres post differentiation[21,36–38]. Moreover, accretional impacts can lead to the complete obliteration of any left-over protoplanetary atmospheres[39]. Therefore, N retained only in their non-atmospheric reservoirs (mantle + core) would be available during subsequent stages of planetary growth. For protoplanets that accreted extremely reduced material like Mercury and aubrite parent body, silicate MOs were the prominent N-bearing non-atmospheric reservoir (Fig.4b). For protoplanets that underwent differentiation at log$f$O$_2$> IW–4, metallic cores were the dominant N-bearing non-atmospheric reservoir (especially for Moon- to Mars-sized planetary embryos). Metallic cores, unlike MOs, are isolated reservoirs after their formation with minimal scope of volatile loss[3]. Thus, the cores can act as the primary N-delivery source for every subsequent stage of planetary growth via partial or complete emulsification of the impactor's core in the target's MO. In contrast to planetary embryos formed by instantaneous accretion (accretion of planetary embryo-sized bodies before undergoing differentiation via heat released by $^{26}$Al decay; Fig.4)[16,30], growth of planetary embryos from differentiated asteroid-sized bodies results in their MO and core reservoirs being extremely N-depleted (<0.01% and <0.5% of the initial accreted N inventory, respectively) (Fig.5; see Methods). Therefore, depending on whether planetary embryos were formed via instantaneous accretion or collisional growth, similar-sized planetary embryos can have very different volatile distribution in their constituent non-atmospheric reservoirs based on their accretion history. Planetary embryos formed via instantaneous accretion contain substantial amount of N in their silicate and core reservoirs while planetary embryos formed via collisional growth are extremely N depleted (Fig.5). Therefore, depending on their accretion history, Moon- to Mars-sized planetary embryos (which acted as seeds for the growth of larger planets like Earth and Venus via collisional accretion) can have their silicate and core reservoirs lying in the range of being almost volatile-free to volatile-bearing.

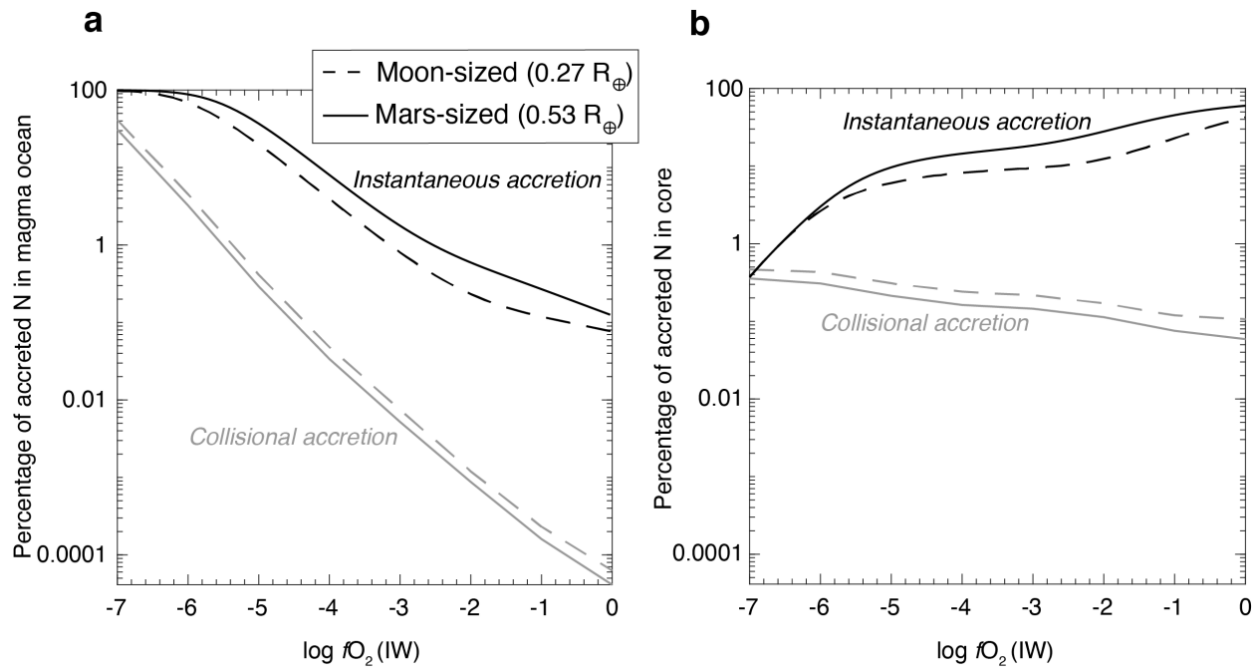

*Figure 5: Comparison of N distribution in non-atmosphere reservoirs of planetary embryos that grew either via instantaneous or collisional accretion.* *Planetary embryos that grew via instantaneous accretion and then underwent differentiation have a much higher percentage of initially accreted N in their (a) magma oceans and (b) cores relative to planetary embryos that grew via collisional accretion of differentiated planetesimals. In this figure, Vesta-sized differentiated bodies (0.04 R$_\oplus$) are assumed to be the seeds of collisional accretion leading to Moon- and Mars-sized planetary embryos.*

**N in the BSE as a constraint of planetary growth regime**

N budget of the present-day planets can be used to test whether planetary embryo growth via instantaneous or collisional accretion was the dominant mechanism in the Solar System. Earth is the only rocky body with constraints of N abundance in its bulk silicate reservoir[1,40]. Although there is a broad consensus that a large rocky planet like Earth had a protracted growth history fueled by the collisional accretion of Moon- to Mars-sized planetary embryos[34,41], the trajectory of its accretionary path is debated. A change in the composition of accreting material from reduced to oxidized as well as oxidized to reduced have been hypothesized to reach the post core-formation log$f$O$_2$ of Earth's primitive mantle, i.e., IW–2 (refs.[34,42]). Assuming protoplanetary core-mantle differentiation at IW–2, in Fig.6a we show the effect of different accretion histories of protoplanets on the final N budget of the BSE (see Methods). Accretion of Earth via planetary embryos that accreted from differentiated asteroids (Scenarios 1-7) would generate a BSE inventory containing 0.001-0.4 PAN (Fig.6a; PAN = Present Atmospheric N content for Earth). A superchondritic C/N ratio of the BSE[3,4,10,13] as well as its N isotopic heterogeneity ($^{15}$N-poor nature of Earth's mantle relative to its atmosphere)[43] suggest that late accretion (post main stage of Earth's growth) of any given class of primitive chondrites cannot explain the present-day N inventory of the BSE post-accretion of extremely N-depleted planetary embryos (Scenarios 1-7). However, growth of Earth via planetary embryos that underwent instantaneous accretion followed by differentiation (Scenarios 8-10) can satisfy the BSE's present-day N budget for 50-1000 ppm of accreted N (Fig.6a). For planetary embryos undergoing differentiation at IW–2, the cores contain most of their non-atmospheric N inventory (Fig.4c); therefore, almost all of the N (>95%) in the target's MO is supplied via emulsification of the impactor's core in the target's MO (see Methods). Efficacy of the retention of atmosphere formed after the last accretion event on Earth (Extended Data Fig.8a) and variation in the percentage of emulsification of impacting planetary embryo's core in the proto-Earth's MO (Extended Data Fig.8b) only affect the exact amount of accreted N required to satisfy the N content of the present-day BSE without affecting the broader conclusions.

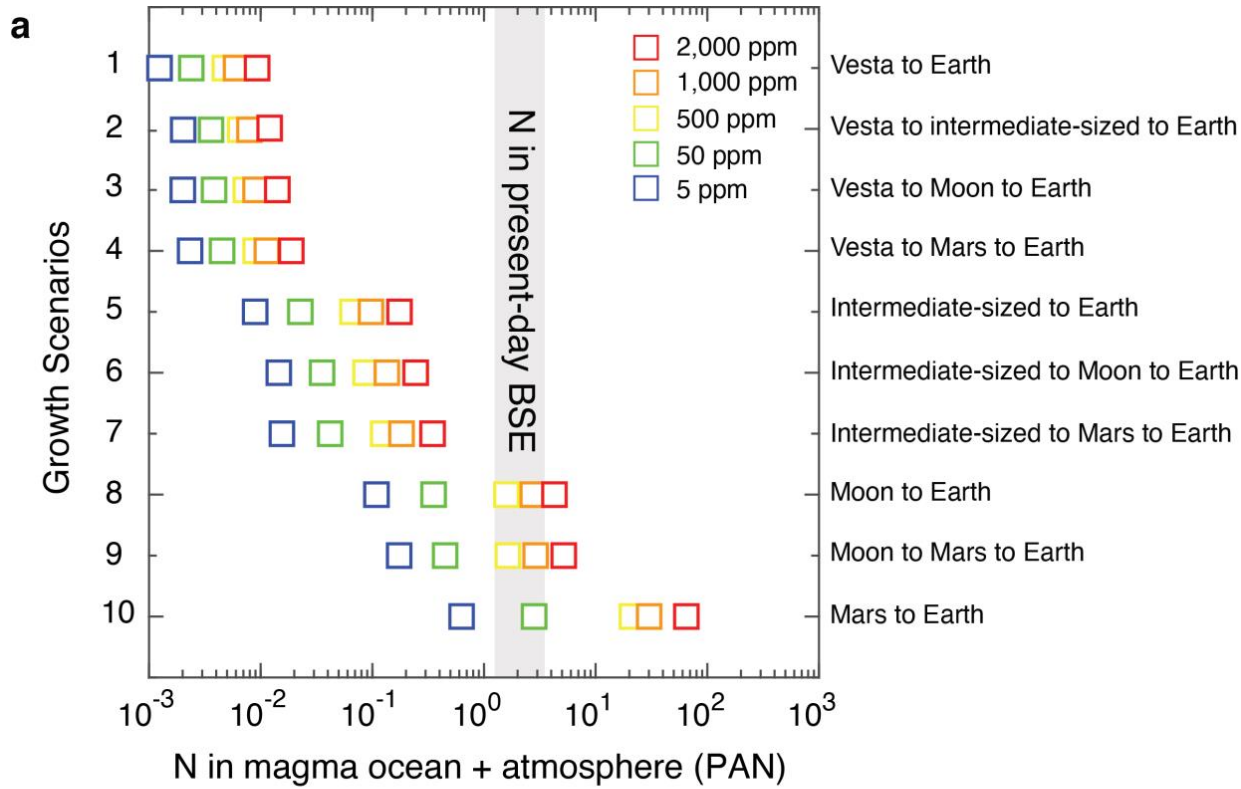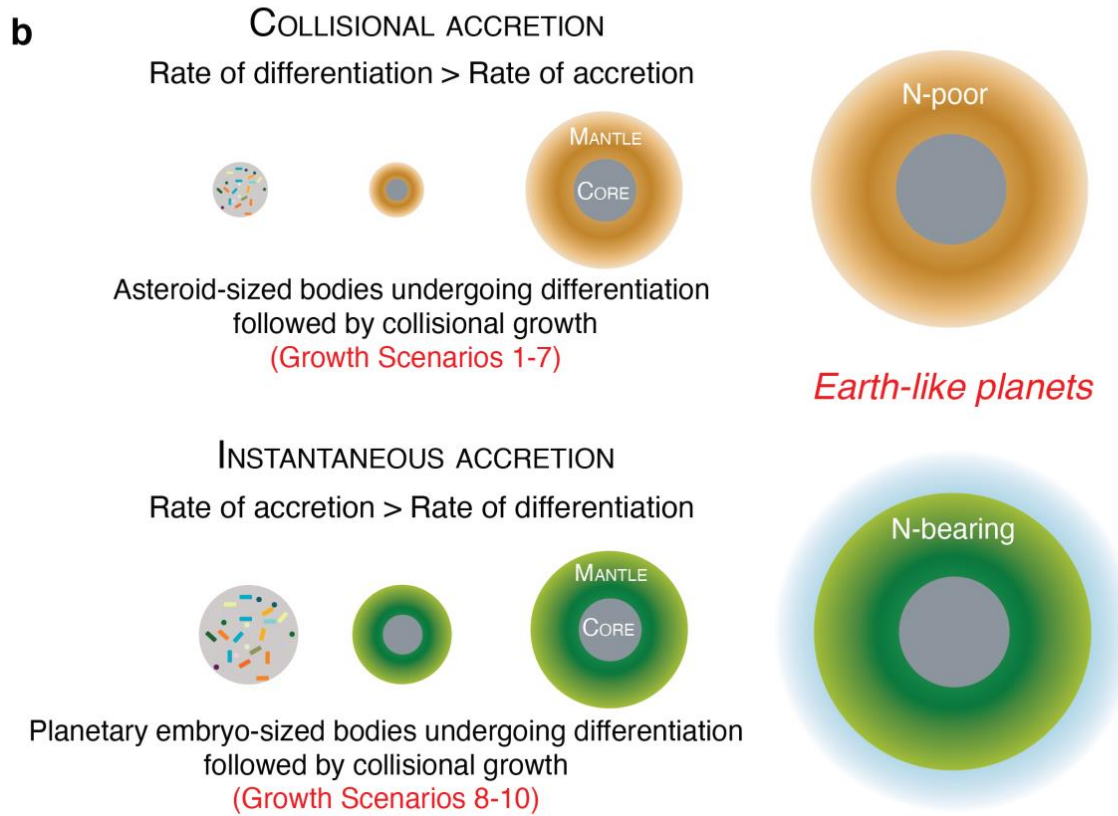

*Figure 6: Effect of rate of protoplanetary accretion versus differentiation on the nitrogen budget of large Earth-like planets. If the rate of differentiation was greater than rate of accretion, then differentiated asteroid-sized bodies would be the primary building blocks for subsequent stages of growth of larger planets like Earth. In such cases (scenarios 1-7), N budget of the BSE cannot be satisfied during the main stage of planetary growth because the mantles and cores of differentiated asteroids were extremely N-poor. However, if the rate of accretion was greater than rate of differentiation, then N budget of the BSE can be set during the main growth period of Earth via primitive rocky bodies which grew to planetary-embryo-size before undergoing differentiation (scenarios 8-10). The highest value of accreted N (2000 ppm) in a) represents the average N content of volatile-rich CI chondrites[5]. N abundance in magma ocean + atmosphere in a) is calculated in terms of the present-day atmospheric nitrogen (PAN) inventory. Right axis label in a) represents end-member growth scenarios of Earth via collisional growth. For example, growth scenario 1 represents Vesta-sized differentiated bodies (0.04 $R_\oplus$; 0.00004 $M_\oplus$,) accreting together to form an Earth-sized planet and growth scenario 2 represents Vesta-sized differentiated bodies (0.04 $R_\oplus$) initially accreting to form intermediate-sized rocky bodies (0.12 $R_\oplus$; 0.001 $M_\oplus$,) which subsequently accrete to form an Earth-sized planet. Grey shaded region represents the estimated N content of the present-day BSE[1,40].*

If the present-day N budget of the BSE was set during the main growth stage of Earth, then the rate of accretion of Moon- to Mars-sized planetary embryos was greater than their rate of differentiation such that planetary embryo-sized protoplanets accreted extremely early, i.e., within decay timescales of $^{26}$Al (Fig.6b). Rapid accretion of planetary embryos followed by their differentiation may deplete N to an extent such that enough N is still available to satisfy the N-budget of large rocky planets during later stages of planetary growth via collisional accretion. Rapid growth of planetary embryos is in agreement with pebble accretion models and geochemical estimates of accretion of Moon- to Mars-sized planetary embryos within ~0.1-2 Myr after the formation of CAIs[16,31,32,44]. This means that the origin of N in large Earth-like planets was linked to the growth rates of protoplanets and the amount of N segregated into their cores.

In summary, we show that protoplanetary differentiation can explain the widespread depletion of N in the bulk silicate reservoirs of rocky bodies ranging from asteroids to planetary embryos. Parent body processes rather than nebular processes were responsible for N (and possibly C) depleted character of the bulk silicate reservoirs of rocky bodies in the inner Solar System. A competition between rates of accretion versus rates of differentiation defines the N inventory of bulk planetary embryos, and consequently, larger planets. N budget of larger planets with protracted growth history can be satisfied if they accreted planetary embryos that grew via instantaneous accretion. Because most of the N in those planetary embryos resides in their metallic portions, the cores were the predominant delivery reservoirs for N and other siderophile volatiles like C. Establishing the N budget of the BSE chiefly via the cores of differentiated planetary embryos from inner and outer Solar System reservoirs[9] obviates the need of late accretion of chondritic materials as the mode of N delivery to Earth. Also, a siderophile character of N and C suggests that their accretional pathways in the inner Solar System planets maybe decoupled from that of water which likely accreted from chondritic materials[1,45].

**Acknowledgements**

The manuscript benefited from the constructive criticism of Sami Mikhail. Gelu Costin is acknowledged for his help with the electron microprobe analyses. D.S.G received support from the NASA FINESST grant 80NSSC19K1538. NASA grants 80NSSC18K0828 and 80NSSC18K1314 to R.D. supported the work. D.S.G. acknowledges additional support from a Lodieska Stockbridge Vaughn Fellowship by Rice University.

**Author Contributions**

D.S.G. conceived and developed the central ideas presented in this study. R.D. helped in refining the ideas. D.S.G. and R.D. designed the experiments. T.H. and D.S.G. performed the experiments. A.F. performed the FTIR analyses. D.S.G. analyzed the experiments and developed the models. D.S.G. and R.D. interpreted the data. D.S.G. wrote the first draft of the manuscript. D.S.G. and R.D. contributed to the revisions.

**Additional information**

Supplementary information is available in the online version of the paper. Reprints and permissions information is available at www.nature.com/reprints. Correspondence and requests for materials should be addressed to D.S.G. (dsg10@rice.edu).

**Competing interests**

The authors declare no competing interests.

**Data availability**

The authors declare that the data supporting the findings of this study are available within the article and its Supplementary Information files. All new data associated with this paper will be made publicly available at www.figshare.com (doi:10.6084/m9.figshare.14191079).

## Methods

**Starting materials**

The starting materials were composed of ~70 wt.% silicate and ~30 wt.% alloy mixtures. A TiO$_2$-free synthetic tholeiite basalt mixture (ThB1), identical to the composition of the silicate mixture of two previous alloy-silicate partitioning studies for N[10,11], was used. The silicate starting mixture was made from reagent grade oxides (SiO$_2$, Fe$_2$O$_3$, Al$_2$O$_3$, Cr$_2$O$_3$, MnO, and MgO) and carbonates (CaCO$_3$, Na$_2$CO$_3$, and K$_2$CO$_3$). The oxides and carbonates were ground and mixed under ethanol in an agate mortar for ~1.5 to 2 h. A CO-CO$_2$ Deltech gas mixing furnace was used to fire the starting silicate mixture at 1000 °C and ~FMQ–2 (FMQ refers to the log$f$O$_2$ of Fayalite-Magnetite-Quartz buffer) for 24 hours to devolatilize the mixture and reduce Fe$_2$O$_3$ to FeO. The alloy mixtures were composed of reagent grade Fe and Ni metals. Si$_3$N$_4$ was used as the N source. Variable amounts of Si were added to the starting alloy mixture to simulate increasingly reduced conditions. The alloy mixture was homogenized under ethanol for ~30 minutes in an agate mortar followed by drying in a desiccator for >2 days. Alloy-silicate mixtures were mixed in 30:70 ratio under ethanol in an agate mortar for ~1 h followed by storage in a desiccator for ~1 day.

**High pressure ($P$)-temperature ($T$) experiments**

The experiments were performed in MgO capsules at a fixed pressure ($P$ = 3 GPa) and three different temperatures (1600, 1700 and 1800 °C) for a given $f$O$_2$. All experiments were performed using an end-loaded piston cylinder device at Rice University. The experiments employed a ½-inch BaCO$_3$/crushable MgO assembly with straight-walled graphite heaters following the calibrations and procedures detailed in a previous study[46]. Temperature was monitored and controlled by a Type C (W5%Re/W26%Re) thermocouple. The experiments were pressurized to the target pressure at room temperature before being heated at a rate of 100 °C/min. Each experiment was sintered overnight at ~850 °C before heating to the target temperature. The experiments were brought to room temperature within ~10-20 s by cutting power to the heater. The recovered samples were cut longitudinally using a W-wire saw, mounted in Crystalbond™, ground using 1200-grit sandpaper, and polished using a 0.3-micron alumina slurry on a velvet cloth. Crystalbond™ was removed from the samples by soaking in acetone overnight. The polished samples were analyzed for N and other major and minor elements using an electron microprobe.

**Demonstration of equilibrium**

We performed a time series at 3 GPa and 1600 °C for 0.5, 2, 6, and 12 hours to demonstrate that the N content in the alloy and silicate melts as well as the silicate melt composition had attained steady state in our experimental products. Extended Data Fig. 2 shows minimal variations in N contents of both alloy and silicate melts as well as silicate melt compositions with time for a given starting mixture. This demonstrates that the experiments at 1600 °C had reached equilibrium at less than 0.5 hours. This means that higher temperature experiments (1700 and 1800 °C) are expected to attain equilibrium at even shorter timescales. Because N diffuses fast at high $T$ (ref.[47]) and based on our time series experiments, the experimental durations of 45-180 min was deemed sufficient. These experimental durations are comparable to or longer than previous experimental studies on C alloy-silicate partitioning experiments in MgO capsules[48,49] and N alloy-silicate partitioning experiments in graphite capsules[10–13].

**Texture of quenched products**

All experimental products were composed of either metal blobs embedded in homogenous silicate glass (Fig. 2a) or metal blobs in a matrix of silicate glass pools and quenched dendritic olivine crystals (Fig. 2b). All experiments also had ferropericlase crystals along the capsule wall. Some ferropericlase was also dispersed within the silicate melt or adjacent to the metallic blobs (Fig. 2a, b). Some experiments also showed the presence of euhedral olivine crystal dispersed in the silicate melt (Fig. 2d). As reported in previous studies[12,50], we observed micron-sized quenched dendritic nitrides in the Fe-Ni alloy melt blobs. No bubbles were observed either in the alloy or silicate phase in any of the experimental products, suggesting that no N was exsolved during quenching. This observation is in agreement with previous studies conducted under similar $P$-$T$ conditions in graphite capsules with roughly equivalent N content in the starting mixtures[10,11].

**Analytical Procedure**

**Electron Probe Micro-analysis (EPMA)**

Major and minor element abundances in the alloy and silicate phases were measured using a JEOL JXA8530F Hyperprobe EPMA at the Department of Earth, Environmental, and Planetary Sciences at Rice University. Following several recent studies[10–13], N in the alloy and silicate phases was also measured using EPMA. The silicate phases were analyzed using carbon-coated samples and standards. For the analysis of the alloy phases, samples and standards were freshly aluminum-coated during each session. All elements, except N, in the silicate phase were measured using natural glass standards from Smithsonian Institute and mineral standards from SPI Supply. Characteristic Kα X-ray lines of the elements in the silicate phases were measured using the following standards: Smithsonian glasses - Si, Al, Ca, Mg, Fe, Na, K, and P, rutile – Ti, chromite – Cr, rhodonite – Mn, and pentlandite – Ni. For alloy phase analysis, the following standards were used: laboratory synthesized stoichiometric $Fe_3C$[51] – C, natural magnetite – O, synthetic Fe metal – Fe, synthetic Ni metal – Ni, and synthetic Si metal – Si. Similar to two previous studies[10,11], synthetic boron nitride (BN) was used as a standard for measuring N in the silicate phases and laboratory synthesized iron nitride $(Fe_3N)$[10] was used as a standard to measure N in the alloy phase.

Accounting for the heterogeneity of the quenched products, 20-micron beam size was used. To measure N along with other elements, 15 kV accelerating voltage and 50 nA beam current was used for silicate phase analysis, and 12 kV accelerating voltage and 80 nA beam current was used for alloy phase analysis. These conditions have been deemed optimum to accurately determine N in the alloy and silicate phases from two previous studies[10,11]. N Kα X-ray counts per second (cps) were measured using LDE2 crystal, also following previous studies[10,11].

The counting time for all elements in the silicate phases, except N, was 10 s on peak and 5 s on each background. The counting time for N in the silicate phase was 80 s on peak and 60 s each on upper and lower background, which yielded a detection limit of ~320 ppm. When the measured N for a given silicate phase in the most oxidized experiments was close to or below the detection limit, N was measured with counting times of 150 s on peak and 300 s on each background resulting in a lower detection limit of ~70 ppm.

A counting time of 10 s on peak and 5 s on each lower and upper background was used to analyze all elements in the alloy except N. A counting time of 80 s per peak and 60 s per each background was used to analyze N. Similar to N analysis in the silicate melts, if the measured N was close to the detection limit (~320 ppm), then N analysis on those samples was conducted with 150 s counting time on peak and 300 s on each upper and lower background which lowered the detection limit to ~70 ppm. After every ~30 measurements on samples, the standards were re-analyzed to account for the effect of C deposition on samples and standards during analysis. Measured C concentration in the alloy phase was corrected by accounting for C blank in Fe-metal as well as any C deposition during an analytical session following the protocol described in previous studies[10,11,49].

**EPMA totals for extremely reduced, N-bearing silicate glasses**

All experiments conducted at ≤~IW–2.5 yielded sum totals (sum of wt.% of all oxides and elemental N) greater than 100 wt.% with increasingly reduced experiments having higher deviations from 100 wt.%. Similar observations have been made in two previous studies[11,13] on N partitioning between alloy and silicate melts. The authors in ref.[11] suggested that the issue of anomalously high totals could be circumvented if an andesitic rather than a basaltic glass standard is used for analyzing the silicate glasses of highly reduced experiments because the polymerization of the silicate matrix of highly reduced experiments is similar to that of more polymerized andesitic silicate glasses. Unlike ref.[11] which had their most reduced experiments at ~IW–4, the present study explored an even more reducing range (up to ~IW–7.1) and it was found that the issue of anomalously high totals cannot be resolved either by using a basaltic, andesitic or rhyolitic glass standard. It has been shown that under extremely reduced conditions N replaces O in the silicate melt structure to form Si-N bonds[28,52]; therefore, counting all of Si in the silicate melt as $SiO_2$, overestimates the concentration of $SiO_2$. Assuming all of N is present in the silicate melt structure as Si-N linkage, we re-calculated $SiO_2$ and the totals of oxides in the silicate melt (represented as $SiO_2$-corr. and Total-corr. in Supplementary table 3). However, even after the corrections the EPMA analytical totals remained higher. Similarly high totals in silicate glasses have been observed in a N-free study examining the alloy silicate partitioning behavior of C under similarly reduced conditions[53]. As Si-$SiO_2$ buffer lies in the range of $fO_2$ where these high totals were observed, $SiO_2$ may be partially reduced to SiO in those experiments which might have led to overestimation of $SiO_2$ content of the silicate melt, and consequently high totals. Thus, in Supplementary Table 3 we have also provided the probable $SiO/SiO_2$ ratios needed to attain EPMA analytical totals of 100.

**Fourier transform infrared spectroscopy (FTIR)**

FTIR spectra were obtained by using a Thermo Nicolet Fourier Transform Infrared Spectrometer at the Department of Earth, Environmental and Planetary Sciences of Rice University (Extended Data Fig. 3). The experimental glasses were doubly polished to thickness of 50-250 μm and cleaned with acetone and ethanol before a given analytical session. A digital micrometer (ID-C125EXB Mitutoyo Digimatic Indicator) was used to measure the sample thickness. Liquid nitrogen was used overnight before every analytical session to remove atmospheric gas contamination. Blank backgrounds were collected at the beginning of each spectral analysis. FTIR spectra were collected on at least three to four spots per sample. Each spectrum was obtained in the frequency range of 650 to 4000 $cm^{-1}$ with a resolution of 4 $cm^{-1}$ and 128 scans using a 100×100 μm spot. The final reported spectra represent the averaged values for each sample.

Only one major peak is found at ~3300 $cm^{-1}$ which has been assigned to N-H species in previous studies[28,54]. However, possible O-H peak at ~3550 $cm^{-1}$ is absent which means that the silicate melts are almost anhydrous. Also, area under the curve of N-H peak is small in comparison with the FTIR spectra of reduced, graphite-saturated silicate glasses with distinctly higher abundance of H (See Fig. 2 in ref.[28]). Therefore, dissolution of N as anhydrous $N^{3-}$ is inferred to dominate in the silicate melts of this study[28]. This is confirmed by the observation that even though N content in the silicate melts increases with decrease in $fO_2$, N-H peak area does not increase with decreasing $fO_2$ (Extended Data Fig. 3). It is important to note that there are no detectable peaks of C-species in the silicate melts in comparison with complex C-O-N-H speciation in hydrated silicate melts in graphite saturated conditions.

**Estimation of oxygen fugacity**

Oxygen fugacity ($fO_2$) of the experimental products was determined via the co-existence of Fe-rich alloy melt and silicate melt:

$FeO^{silicate\ melt} = Fe^{alloy\ melt} + ½\ O_2$  (Eq. 1)

from which $fO_2$ relative to $fO_2$ of the iron–wüstite buffer (ΔIW), at a given *P-T*, is defined by:

$$\Delta \text{IW} = 2 \log \frac{a_{\text{FeO}}^{\text{silicate melt}}}{a_{\text{Fe}}^{\text{alloy melt}}} = 2 \log \frac{X_{\text{FeO}}^{\text{silicate melt}} \gamma_{\text{FeO}}^{\text{silicate melt}}}{X_{\text{Fe}}^{\text{alloy melt}} \gamma_{\text{Fe}}^{\text{alloy melt}}} \quad \text{(Eq. 2)}$$

where, $a_{\text{FeO}}^{\text{silicate melt}}$ is the activity of FeO component in silicate melt and $a_{\text{Fe}}^{\text{alloy melt}}$ is the activity of Fe component in alloy melt; $X_{\text{FeO}}^{\text{silicate melt}}$ and $\gamma_{\text{FeO}}^{\text{silicate melt}}$ is the mole fraction and activity coefficient of FeO component in silicate melt, respectively; $X_{\text{Fe}}^{\text{alloy melt}}$ and $\gamma_{\text{Fe}}^{\text{alloy melt}}$ is the mole fraction and activity coefficient of Fe component in alloy melt, respectively. Using the non-ideal solution model, $fO_2$ was calculated assuming a fixed $\gamma_{\text{FeO}}^{\text{silicate melt}}$ of 1.5 (ref.[55]). To account for the non-ideal interactions between components of the alloy melt, $\gamma_{\text{Fe}}^{\text{alloy melt}}$ was calculated via ε approach in Wagner equations[56] using the 'Online Metal Activity Calculator' (http://norris.org.au/expet/metalact/).

**Comparison with the graphite-undersaturated data of Roskosz et al. (2013)**

In a previous study[15], at 5-10 GPa and a limited $fO_2$ range (between IW–2.7 and IW–1.5), order of magnitude higher $D_N^{\text{alloy/silicate}}$ values were reported from laser heated diamond anvil cell (LHDAC) experiments at graphite-undersaturated conditions in comparison to graphite-saturated systems conducted using multi anvil[15]. However, the quantitative effects of C content in alloy on $D_N^{\text{alloy/silicate}}$ could not be determined from that study owing to the unavailability of C content measurements of their alloys. Furthermore, it also remained unknown whether the effect of C content of alloy on $D_N^{\text{alloy/silicate}}$ persists across the entire $fO_2$ range (between IW–7 and IW–1)[33] applicable for protoplanetary and planetary differentiation.

**Parametrization of $D_N^{\text{alloy/silicate}}$**

To predict alloy-silicate partitioning behavior of N for a wide range of graphite-undersaturated MO settings, we derived an empirical parametrization for $D_N^{\text{alloy/silicate}}$ using data from this study and previous studies[10–13,15,25,29] by incorporating an additional term for the effect of variation of C content in the alloy. We note that this is the first $D_N^{\text{alloy/silicate}}$ parameterization, which can be applied for both graphite-undersaturated and graphite-saturated alloys. 164 experiments used for parametrization comprised the range: *P* = 1 to 17.7 GPa, *T* = 1400 to 2577

°C, log$fO_2$ = IW–7.1 to IW–0.2, S content of alloy = 0 to 32.1 wt.%, Si content of alloy = 0 to 24.6 wt.%, NBO/T = 0.4 to 2.5 and C content of alloy = 0.1 wt.% to graphite-saturated. The empirical equation is based on similar thermodynamic considerations as detailed in ref.[11].

$$\ln D_N^{alloy/silicate} = a + \frac{b}{T} + c\frac{P}{T} + d \ln(100 - X_S^{alloy}) + e \ln(100 - X_S^{alloy})^2 + f \ln(100 - X_{Si}^{alloy})^2 + g \ln(100 - X_C^{alloy}) + h \ln(100 - X_C^{alloy})^2 + i\,NBO/T + j \ln X_{FeO}^{silicate} \quad (Eq.\ 3)$$

where, $P$ is pressure in GPa, $T$ is temperature in K, $X_{FeO}^{silicate}$ is the FeO content of the silicate melt in wt.% and NBO/T term accounts for the effect of silicate melt composition. The presence of S, Si, and C in the alloy melt was accounted by the terms: $\ln(100 - X_S^{alloy})$, $\ln(100 - X_{Si}^{alloy})$, and $\ln(100 - X_C^{alloy})$, where $X_S^{alloy}$, $X_{Si}^{alloy}$, and $X_C^{alloy}$, denote wt.% of S, Si, and C in the alloy melt, respectively. A built-in 'regress' function in Matlab® was used to perform regression with an unweighted least squares minimization scheme. The resulting coefficients and their 1-σ uncertainties are shown in Supplementary Table 5. The experimental and predicted data show a good agreement (Extended Data Figure 7a). The experiments from ref.[12] containing high initial TiO$_2$ were not included in the parametrization because they yielded substantially higher $D_N^{alloy/silicate}$ values at log$fO_2$<~IW–3 in comparison to the data of refs.[11,13] primarily due to the formation of osbornite (TiN) speckles attached to the metallic alloys[12]. The authors of that study[12] had also noted the high $D_N^{alloy/silicate}$ values yielded by those experiments "are an experimental artifact not representing true partitioning values". Two experiments (HB01 and HB15) from ref.[25] were not included in the parametrization because of the low analytical totals (<95 wt.%) and anomalously low C contents of their alloy phases relative to similar studies conducted under graphite-saturated conditions.

Extended Data Figure 7b shows the comparison of the empirical $D_N^{alloy/silicate}$ parametrization developed in this study for the entire range of C content in the alloy (C-free alloys to graphite-saturated) with the $D_N^{alloy/silicate}$ parametrizations of two previous studies determined in graphite-saturated conditions[11,12]. Predicted $D_N^{alloy/silicate}$ values for graphite-saturated alloys by the parameterization of this study is within the range of predicted $D_N^{alloy/silicate}$ of ref.[11] at log$fO_2$> ~IW–4. Below ~IW–4, the predicted $D_N^{alloy/silicate}$ values of ref.[11] decrease sharply due to the incorporation of even small amounts of Si (<0.5 wt.%) in graphite-saturated alloys, which is not captured by the experiments and parameterization of this study. $D_N^{alloy/silicate}$ values predicted in graphite-saturated alloys by the parametrization of this study are almost similar to those of ref.[12] across the entire $fO_2$ range. The predicted $D_N^{alloy/silicate}$ values for C-free alloys as well as C-poor alloys (0.4 wt.%; concentration similar to the estimated C content of Earth's core[4]), in agreement with the experimental dataset of this study (Fig. 2b), are approximately an order of magnitude higher than those for systems with graphite-saturated alloys at log$fO_2$> ~IW–4. Below IW–4, the gap between the predicted $D_N^{alloy/silicate}$ values in graphite-undersaturated and graphite-saturated conditions diminishes because C solubility in the alloy melt drops with the incorporation of Si under such conditions[11,53]. Below IW–6, the predicted $D_N^{alloy/silicate}$ values in graphite-undersaturated and graphite-saturated alloy bearing systems are almost similar because C solubility in the alloy melt approaches the assumed C content (0.4 wt.%) of graphite-undersaturated alloys.

**Numerical Modeling**

**N re-distribution between silicate magma ocean, metallic core, and atmosphere**

Post large-scale melting, the accreted N is distributed between three major reservoirs in a rocky body – atmosphere, silicate magma ocean (MO), and alloy core. The atmosphere overlaying the MO sets N abundance in the silicate melts via its vapor pressure induced solubility. The equilibration of the alloy melt with the MO determines the N inventory of the core of the rocky body.

Using a mass balance, the total mass of N is conserved pre- and post-differentiation:
$$M_N^{tot} = M_N^{atm} + M_N^{MO} + M_N^{core}$$
where $M_N^{tot}$, $M_N^{atm}$, $M_N^{MO}$, and $M_N^{core}$ represent the mass of N in the bulk body, atmosphere, magma ocean, and core, respectively.

Concentration of N in a given reservoir is represented by:
$$C_N^i = M_N^i/m^i$$
where $C_N^i$ represents the concentration of N in a given reservoir and $m^i$ represents the mass of that reservoir.

Equilibrium between the alloy and silicate is given by $D_N^{alloy/silicate} = C_N^{alloy}/C_N^{silicate}$. $D_N^{alloy/silicate}$ is calculated as a function of $fO_2$ using Eq. 3 of this study. Instead of using a fixed

Henry Law's constant to calculate $C_N^{silicate}$, as done in previous studies[3,4,10], in this study $C_N^{silicate}$ is calculated as a function of $fO_2$ using the following equation from ref.[24] ($C_N^{silicate}$ = 0.06 $p_N$ + 5.97 $p_N^{1/2}$ $fO_2^{-3/4}$) (Extended Data Fig. 9). The partial pressure of N is given by $p_N$ = $M_N^{atm}$.g/A, where g and A is the gravitational constant and area of a rocky body or MO surface (Supplementary table 7).

**Collisional growth of asteroid- and planetary embryo-sized rocky bodies**

For collisional growth, the rocky bodies are assumed to grow by accretion of a fixed ratio of the resultant rocky body's mass (=seed rocky body's mass/final rocky body's mass) at each stage of accretion. The mass of the target increases via collisional accretion of the impactor (seed rocky body) at every step of growth. For all scenarios of rocky body accretion, each of the impactors is assumed to have been differentiated at a given $fO_2$ with N distribution in their resulting reservoirs calculated via coupled metal-silicate-atmosphere fractionation model. At every step of collisional accretion, we assume: 1) global scale melting of the target's and impactor's mantles[34,57], 2) complete emulsification of impactor's core in target's MO[34,57], and 3) complete loss of atmospheric reservoirs of both the target and the impactor pre-collision or/and during the impact[38,39]. The total amount of N present in the mantle of the target and the impactor as well as N present in the impactor's core (assumed to be released into the target's MO post-impact) contribute to the post-impact N budget available for metal-silicate-atmosphere fractionation. The core of the target is assumed to be a non-interacting, isolated reservoir. Post-impact, the net silicate MO (=mass of target's + impactor's mantle) equilibrates with the MO degassed atmosphere (dependent on the vapor pressure induced solubility). N present in the MO is available to equilibrate with the impactor's core forming alloy (dependent on $D_N^{alloy/silicate}$). N exchange between all three reservoirs is calculated simultaneously by the coupled metal-silicate-atmosphere equilibration model as defined above. Finally, the core-forming metal sinks down to form the net post-merger core. We assume that N content of the silicate mantle remains unmodified as it crystallizes from the MO stage to form the solid mantle. N in the silicate portion of the post-merger body is available for exchange when it acts as a target for the next stage of collisional growth, while N in its core would remain isolated and N in its atmosphere is lost either during collision or due to the rocky body's inability to retain its resulting atmosphere (applicable to asteroids and small planetary embryos).

**Collisional growth of Earth**

The growth scenarios are categorized as ten end-member scenarios based on the size of the body that originally underwent differentiation and later collided to form larger bodies with increase in MO depth at every step of accretion: 1) Scenarios 1-4: Vesta-sized bodies (0.04 $R_\oplus$; 0.00004 $M_\oplus$), 2) Scenarios 5-7: Intermediate-sized bodies (0.12 $R_\oplus$; 0.001 $M_\oplus$), 3) Scenarios 8-9: Moon-sized bodies (0.27 $R_\oplus$; 0.012 $M_\oplus$), and 4) Scenario 10: Mars-sized bodies (0.53 $R_\oplus$; 0.107 $M_\oplus$); where $R_\oplus$ and $M_\oplus$ is the radius and mass of the present-day Earth, respectively. The collisional growth of Earth follows a similar framework as explained in the previous section. Growth of Earth via accretion of asteroid-sized bodies acting as seeds (Scenarios 1-7) would lead to an extremely N-depleted BSE, while growth of Earth via planetary embryo-sized bodies acting as seeds (Scenarios 8-10) can satisfy the N budget of the BSE for varying values of net accreted N in the seed planetary embryo (Fig. 6). As described in the previous section, N present in the MO of the target and the impactor as well as N in the impactor's core contribute to the net N available for metal-silicate-atmosphere fractionation. The atmosphere of the target and the impactor are both assumed to be lost either pre-collision or/and stripped away during every stage of collision. However, after the final accretion event on Earth, i.e., when Earth has attained almost all of its mass (possibly after the Moon formation event), the final atmosphere formed after the MO degassing is assumed to be retained in the calculations presented in Figure 6. This is a reasonable assumption as late accretion events after the Moon forming impact may not be energetic enough to lead to the complete removal of Earth's atmosphere. If there were substantial amount of atmospheric loss post last step of accretion, then consistently higher accreted bulk N values would be required. For example, in Scenario 10, for 50 ppm of accreted N in Mars-sized planetary embryos that underwent differentiation, ~60-100 % of atmospheric retention would be enough to satisfy the N budget of the present-day BSE and for lesser degrees of atmospheric retention higher amount of bulk accreted N would be required (Extended Data Fig. 8a).

Also, we assumed complete emulsification of the impactor's core in the target's MO after every stage of collision for the results presented in Figure 6. This is a reasonable assumption for the asteroid-sized bodies as well as Moon-sized planetary embryos[34,57]. However, this assumption may not hold true for relatively large, Mars-sized impactors[58,59]. To account for the effect of inefficient emulsification of impactor's core in target's MO, we provide an additional set of example calculations for Scenario 10 with 50 ppm of accreted N. Greater than 50% emulsification of the impactor's core in the target's MO during every stage of impact would satisfy the N budget of the present-day BSE if Mars-sized impactors had accreted 50 ppm N. With lesser degrees of emulsification of the impactor's core in the target's MO would require higher amount of bulk accreted N (Extended Data Fig. 8b).

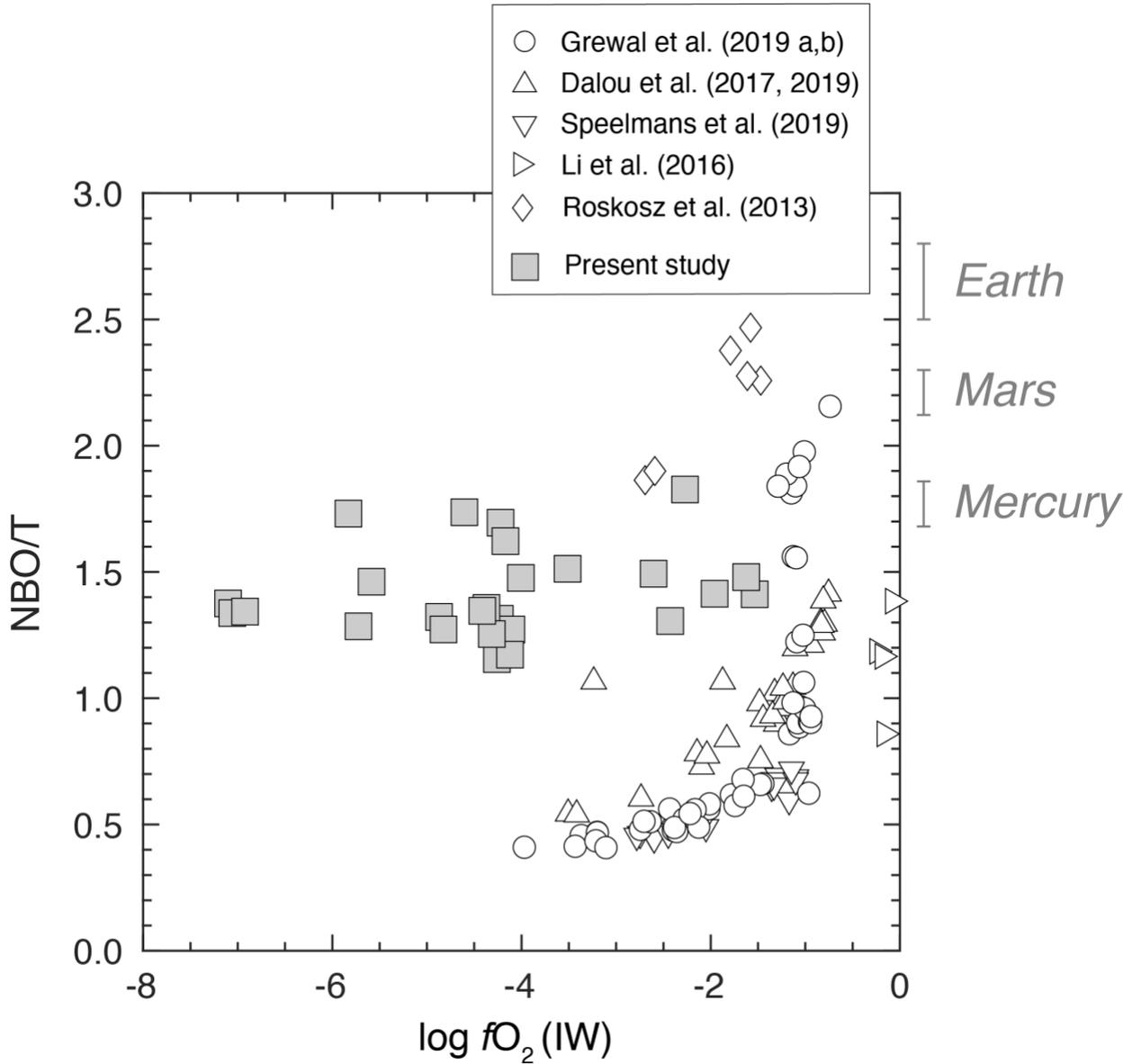

*Extended Data Figure 1: Comparison of the silicate melt compositions of this study with previous studies as a function of oxygen fugacity.* Below IW–1.5 ($fO_2$ range explored in this study), the silicate melt compositions of this study are more mafic, i.e., have higher NBO/Ts, relative to the silicate melt compositions used to determine $D_N^{alloy/silicate}$ in previous studies. Therefore, the silicate melt compositions of this study are more representative of magma oceans of inner Solar System rocky planets. Primitive mantle compositions are used to estimate the magma ocean compositions of Earth[60], Mars[61] and Mercury[62]. NBO/T is a measure of degree of silicate melt polymerization and is expressed as total non-bridging oxygens per tetrahedral

*cations; NBO/T = (2 × Total O)/T – 4, where T = Si + Ti + Al + Cr + P). The calculated error bars for NBO/T represent ±1-σ deviation based on the replicate electron microprobe analyses and are smaller than the symbol sizes for all data from this study.*

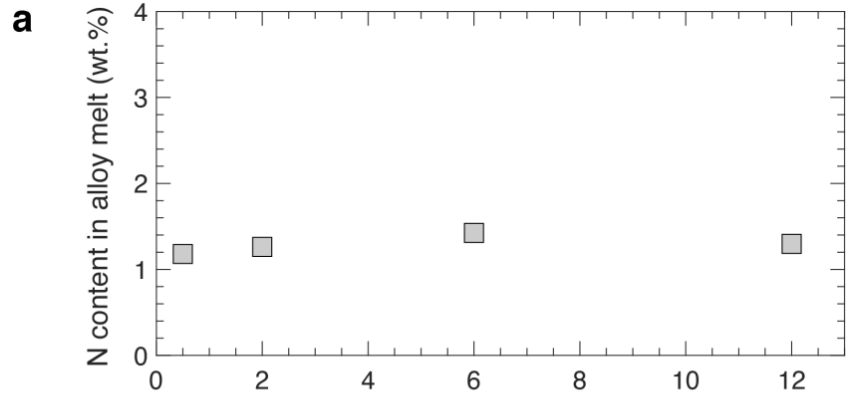
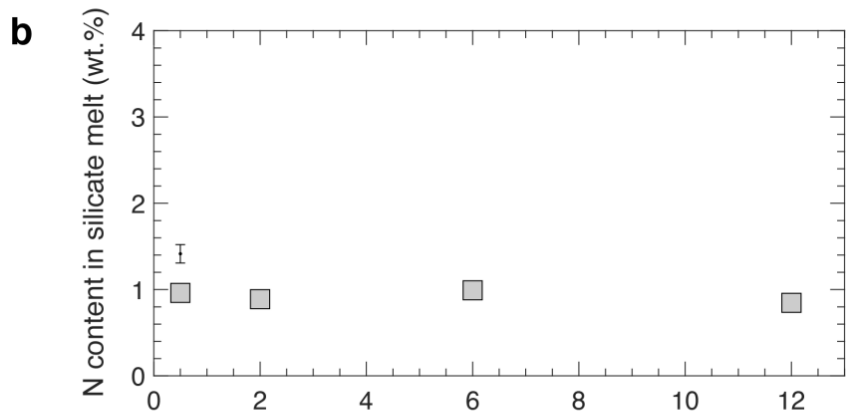
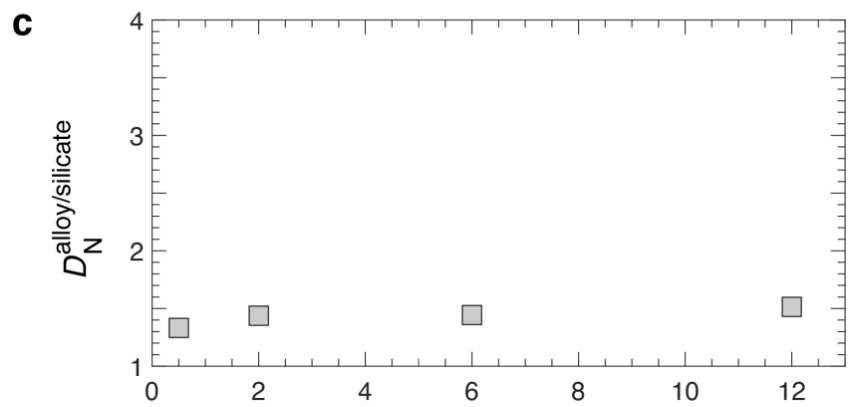
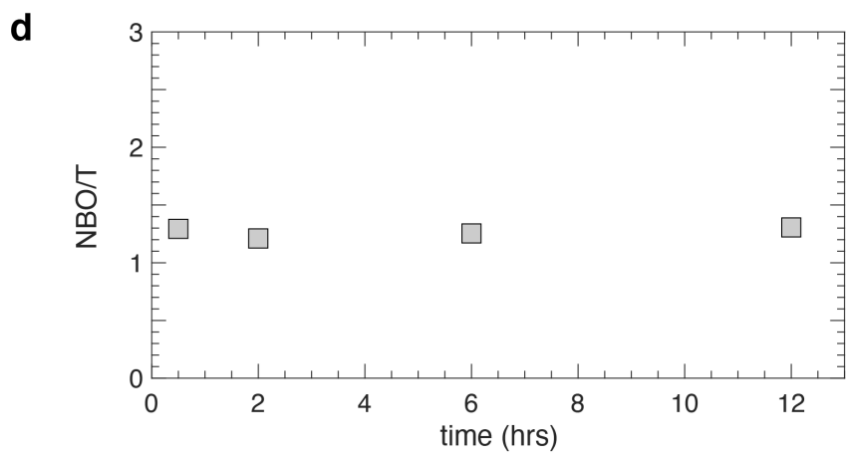

***Extended Data Figure 2: Time series to determine the experimental duration necessary to reach equilibrium.** N contents in the (**a**) alloy and (**b**) silicate melts, and consequently (**c**) $D_N^{alloy/silicate}$ show no variation beyond the uncertainties of the measurements as a function of time for experiments conducted at 3 GPa and 1600 °C for 0.5, 2, 6, and 12 hours (Experiment numbers: X63, G634, X74, and G639). These demonstrate that N exchange between the two phases had attained equilibrium at less than 0.5 hours at our experimental conditions. Also, an almost unchanged N content in alloy and silicate melts with increase in time means that there was no loss of N from the alloy + silicate melt system with increase in experimental run time. (**d**) NBO/T of the silicate melt compositions also show no variation with time beyond the uncertainties of the measurements, which illustrates that the silicate melt compositions had also reached steady state. All experiments were conducted with a fixed starting composition of alloy + silicate mixture (70%ThB1+30%Fe-5Ni-5N-17.5Si). Error bars in all panels are ±1-σ deviation based on replicate electron microprobe analyses and where absent the error bars are smaller than the symbol size.*

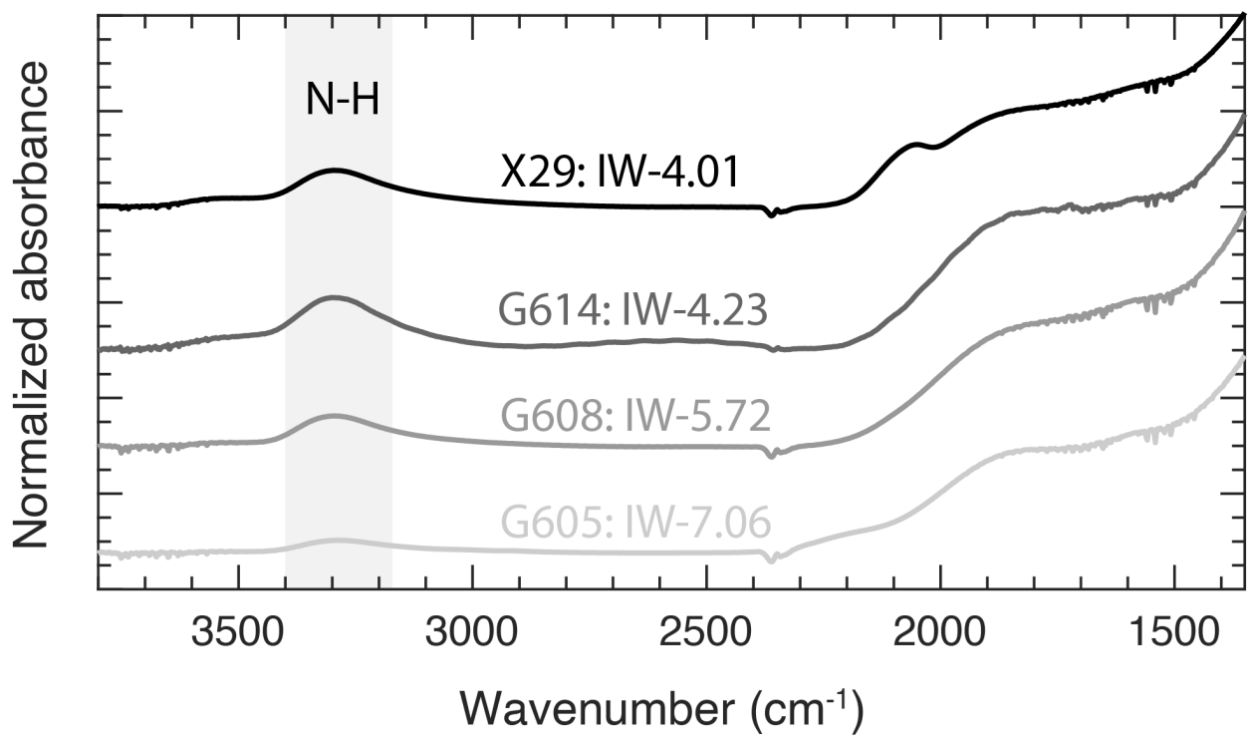

*Extended Data Figure 3: Thickness normalized FTIR spectra of the peaks associated with nitrogen species in the experimental silicate glasses of this study.* The only detectable, IR-active N-bearing peak was that of N-H stretching, marked by the grey band. No C-species were detectable.

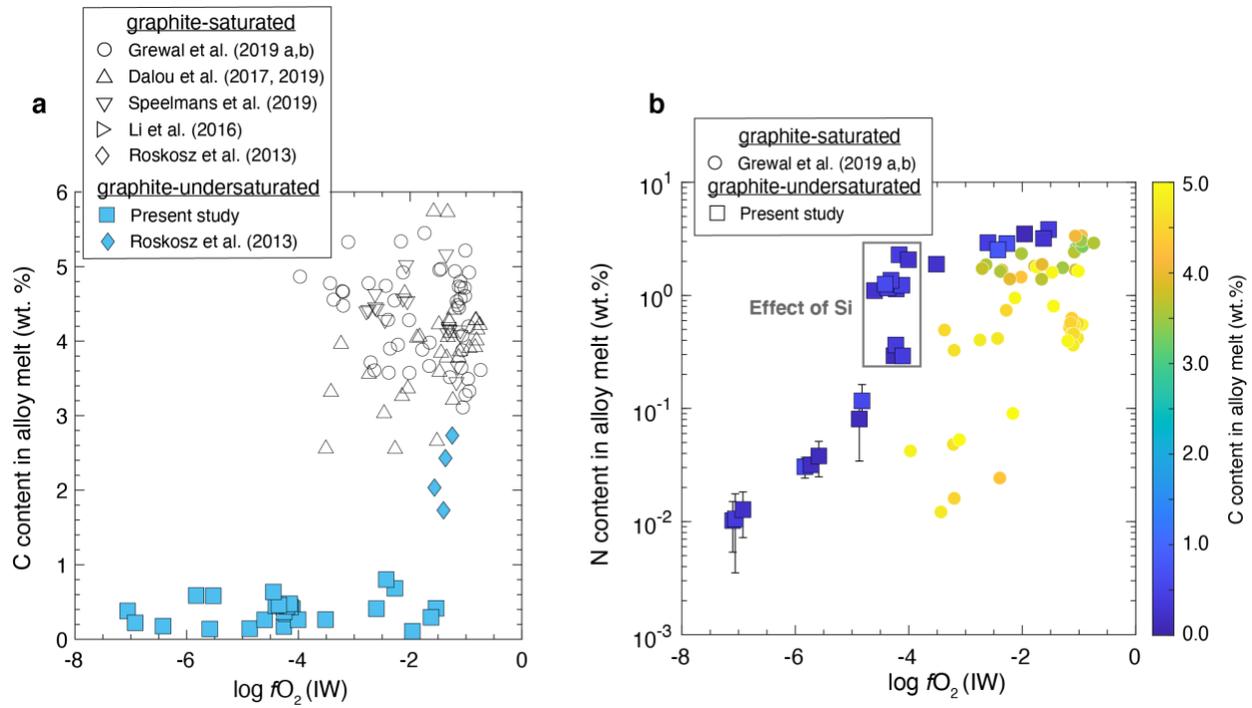

*Extended Data Figure 4: Carbon and nitrogen contents in the alloy melts as a function of oxygen fugacity. a)* C content in the alloy melt in graphite-undersaturated experiments of this study is substantially lower (0.11-0.80 wt.%) than the graphite-saturated experiments of the previous studies[10–13,25,29]. *b)* In agreement with previous studies in graphite-saturated conditions, N content in the alloy melt decreases with decrease in $fO_2$ in graphite-undersaturated conditions. However, at any given $fO_2$, N in graphite-undersaturated alloys is substantially greater than graphite-saturated alloys. $D_N^{alloy/silicate}$ for graphite-saturated alloys has been determined only in N-undersaturated conditions, therefore, N content in the alloys from only two previous studies[10,11] was compared with the data from the present study because of similar N contents in the starting mixtures. Error bars represent ±1-σ deviation based on the replicate electron microprobe analyses; where absent, the error bars are smaller than the symbol size.

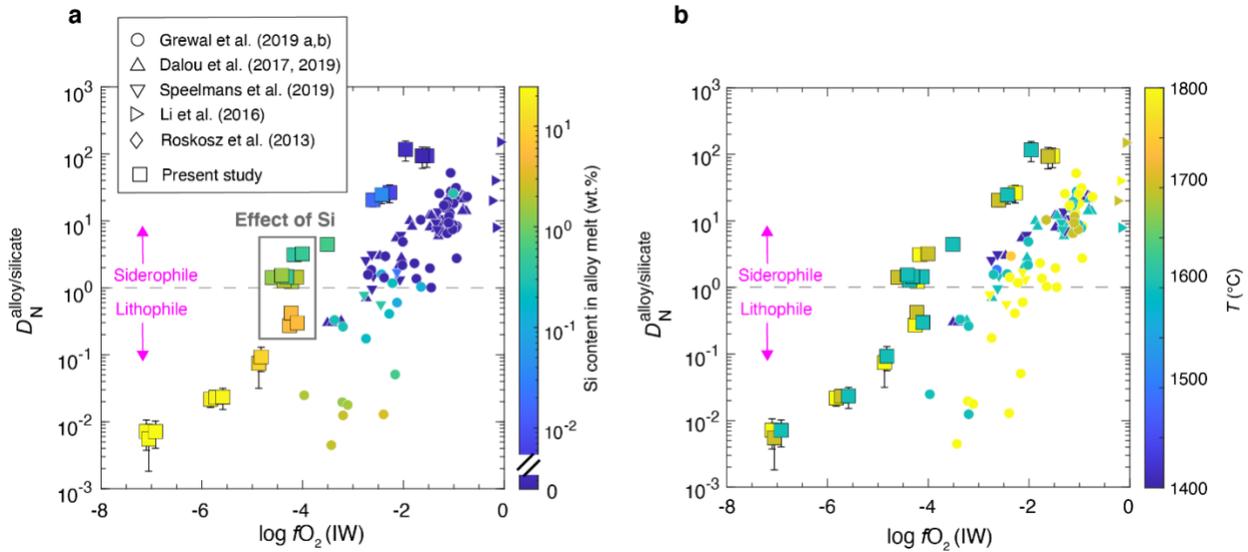

***Extended Data Figure 5:** $D_N^{alloy/silicate}$ as a function of oxygen fugacity and (a) silicon content in the alloy and (b) temperature. **a)** In addition to the effect of fO$_2$, when Si content in the alloy melt and FeO content in the silicate melt are coupled to each other, incorporation of Si into the alloy melt on its own has a strong negative effect on N content in the alloy at a similar logfO$_2$ (~IW–4; here shown by grey rectangle). A similar effect has been observed in a previous study[11] albeit at a higher logfO$_2$ (~IW–2.5) because in graphite-saturated alloys, Si expels N from the alloy melt even at lower concentrations (as low as 0.1 wt.% Si). **b)** In contrast to the observations of previous studies[11,12] in graphite-saturated conditions, temperature does not have any discernible effect on $D_N^{alloy/silicate}$ in the limited temperature range explored in this study. Error bars represent ±1-σ deviation obtained by propagation of ±1-σ deviation on N content in the alloy and silicate melts; where absent, the error bars are smaller than the symbol size.*

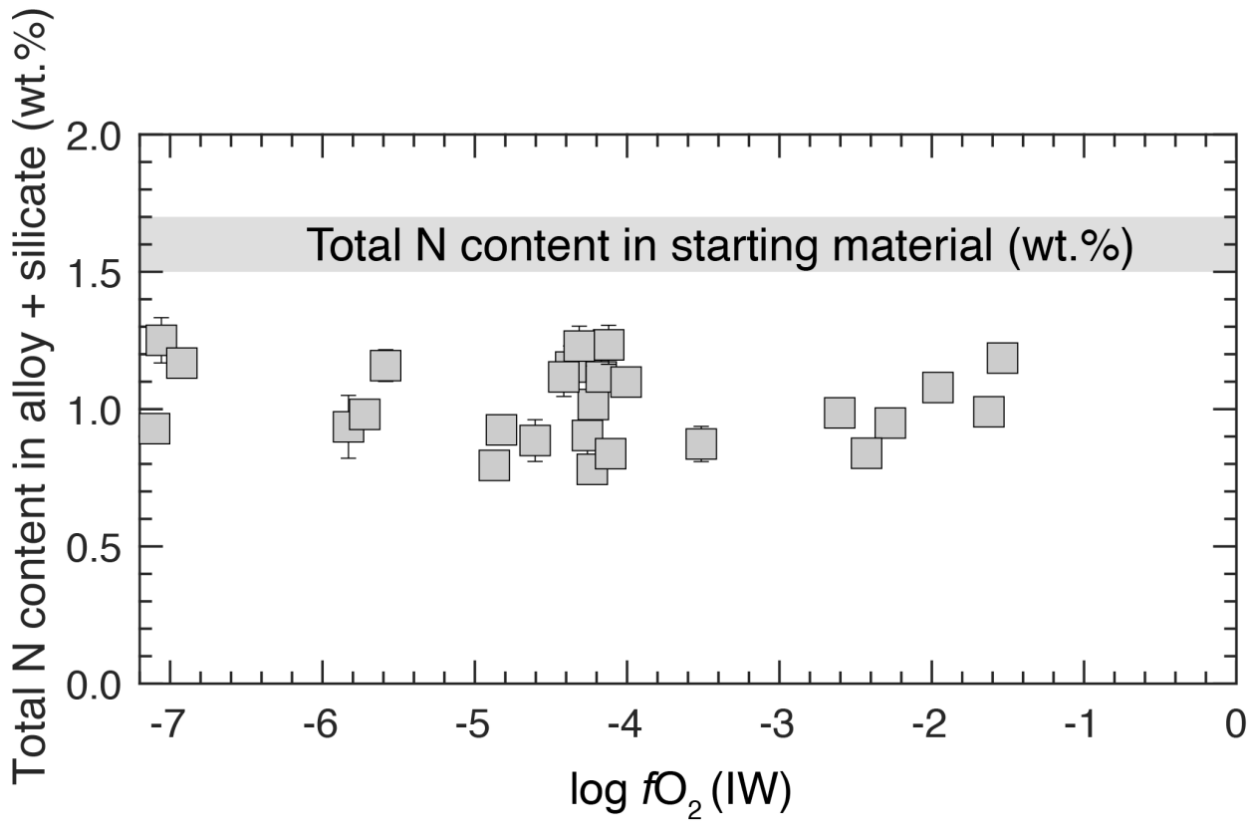

*Extended Data Figure 6: Comparison between nitrogen content in the final products and starting mixtures.* Similar to the observation in all previous studies that estimated $D_N^{alloy/silicate}$ in graphite capsules[10–13,25,29], the reconstructed N content in the final products of this study in MgO capsules is less than the N content in the starting mixture. Mass balance suggests that the extent of recovery of initial N content lies in the range of ~50-85%. Loss of N has been explained by storage of N in the pores of the capsule walls or diffusive loss across the capsule wall[11,12]. Error bars represent ±1-σ deviation obtained by propagation of ±1-σ deviation on N content in the alloy and silicate melts; where absent, the error bars are smaller than the symbol size.

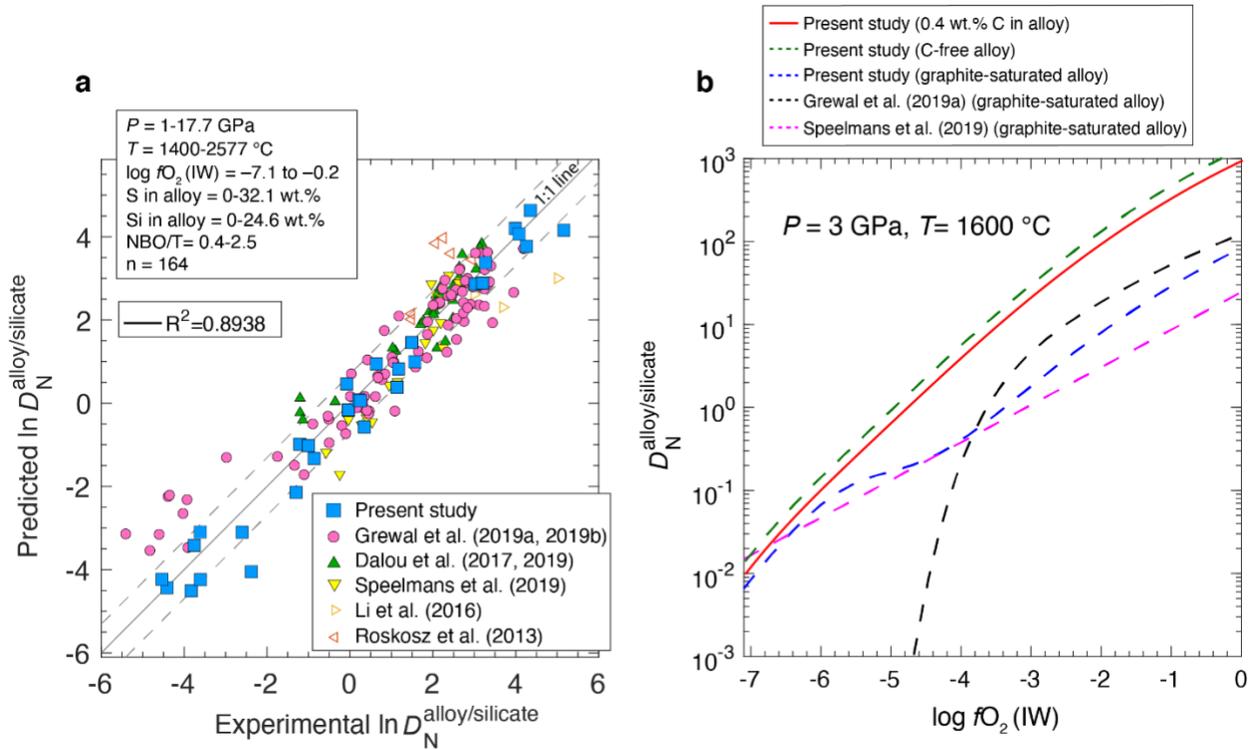

*Extended Data Figure 7: Predicted $D_N^{alloy/silicate}$ using the parametrization developed in this study and comparison between the $D_N^{alloy/silicate}$ values predicted by this study and two previous studies. a) Predicted $D_N^{alloy/silicate}$ using the parametrization developed in this study plotted against experimentally determined $D_N^{alloy/silicate}$ for Fe-Ni-N±C±S±Si alloy melt-silicate melt equilibration. 'n' represents the total number of experiments that were used to calibrate the parameterized equation in this study. Solid line represents 1:1 fit while the dashed lines represent error within a factor of 2. b) The predicted $D_N^{alloy/silicate}$ values in C-free and graphite-undersaturated alloys, in agreement with the experimental data of this study, are an order of magnitude higher than the predicted $D_N^{alloy/silicate}$ values of graphite-saturated alloys at logfO$_2$> IW–5 and the gap between the predicted values decreases with decrease in fO$_2$ (see Methods).*

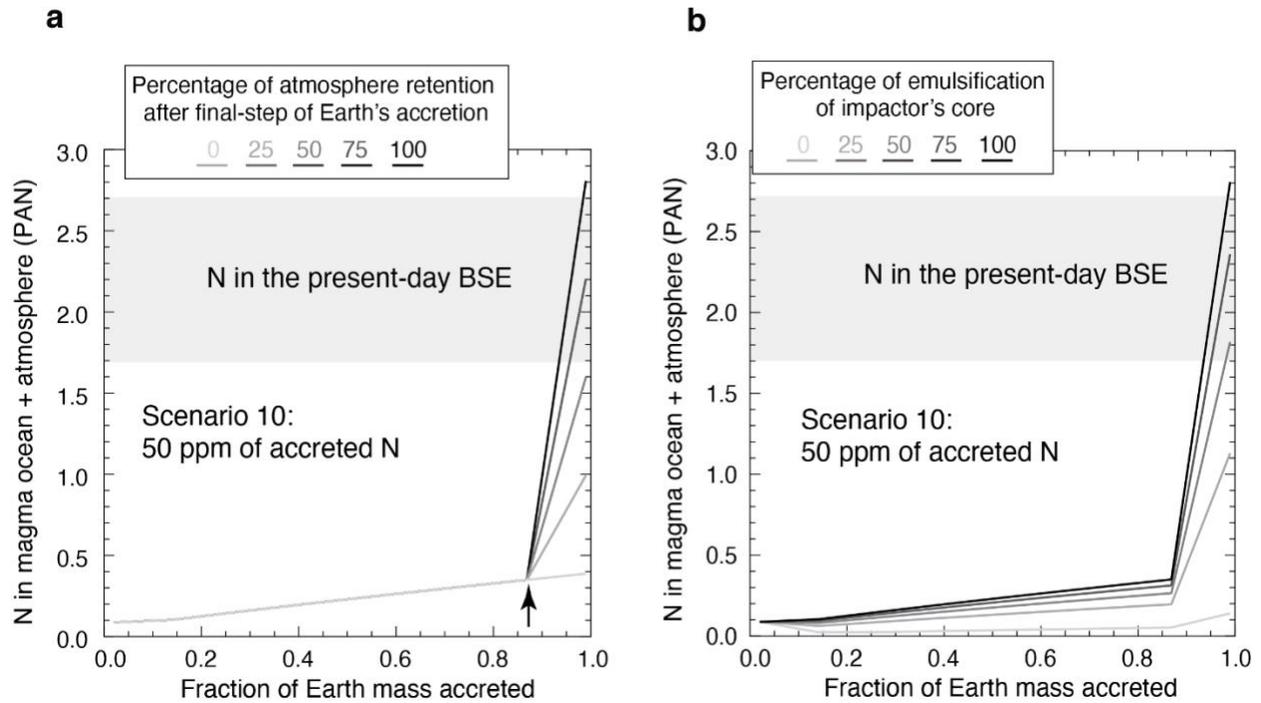

*Extended Data Figure 8: The effects of (a) the extent of atmosphere retention after final-step of Earth's accretion and (b) the extent of emulsification of the impactor's core in the target's magma ocean on N budget of the BSE.* For scenario 10 (as defined in Fig. 5a) and 50 ppm of accreted N, it can be seen that N budget of the present-day BSE can be satisfied for ~60-100 % of atmosphere retention on Earth after its final accretion event (a) and for ~50-100 % emulsification of the impactor's core in the target's MO (b) during every step of accretion. Lesser extent of final-stage atmospheric retention or lesser degree of emulsification of the impactor's core would require higher amount of accreted N (>50 ppm) in the seed planetary embryos (here Mars-sized) to satisfy the N budget of the present-day BSE.

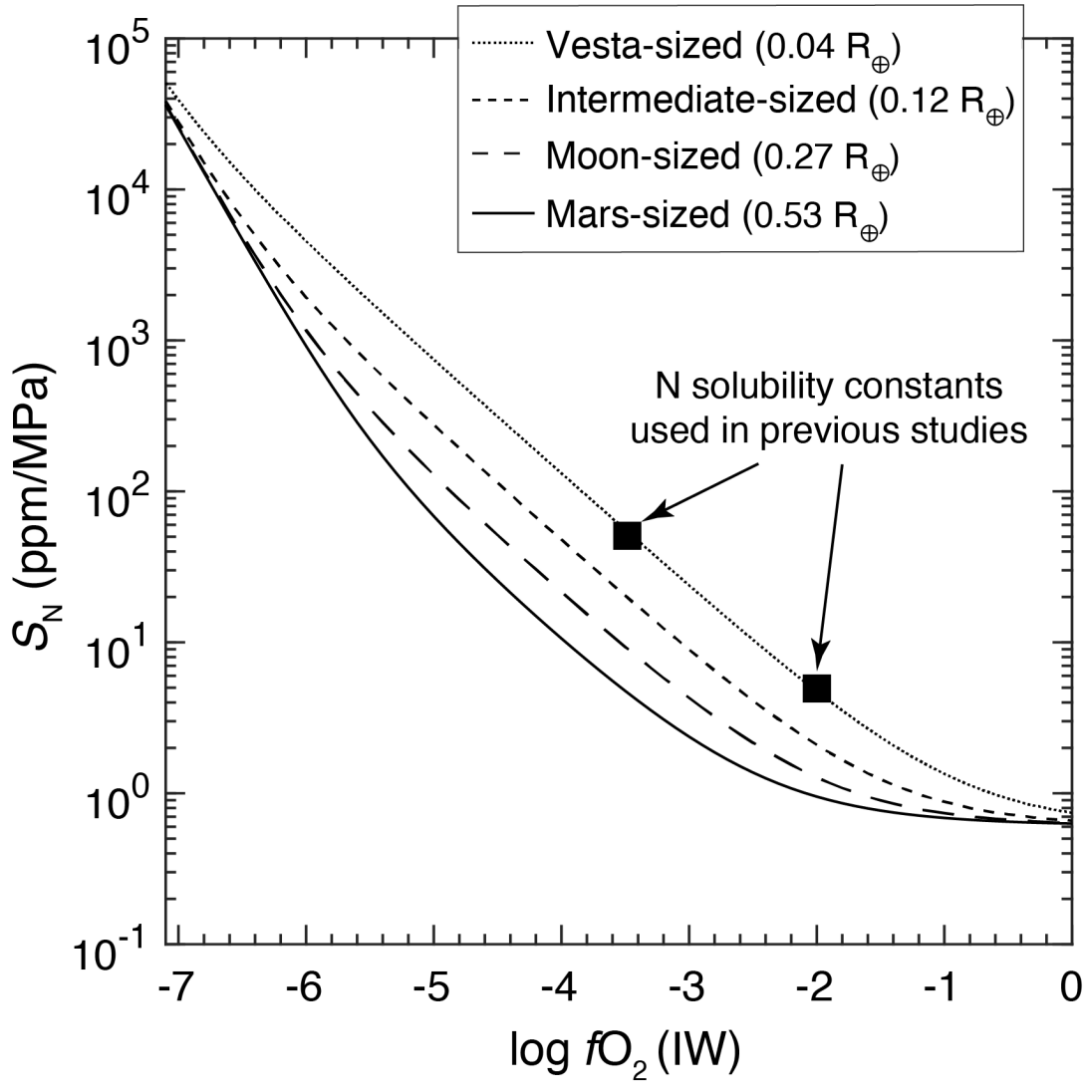

*Extended Data Figure 9: Comparison between the calculated nitrogen solubility constants ($S_N$) from this study and the fixed values used in previous studies. The effective solubility constants for N vary with the size of protoplanetary bodies with variations in the range of an order of magnitude from a Vesta-sized to a Mars-sized protoplanet. Therefore, using a fixed solubility constant, as used in previous studies[3,4,10] to calculate the solubilities of N in magma oceans for rocky bodies having different sizes can give erroneous results.*